\documentclass[twocolumn,prl,amsmath,amssymb,superscriptaddress,notitlepage,footinbib]{revtex4-1}

\usepackage{amsmath,amsfonts,bm}
\usepackage[margin = 0.75in]{geometry}
\usepackage{fancyhdr}
\usepackage{braket}
\usepackage{physics}
\usepackage{comment}
\usepackage{dsfont}
\usepackage{graphicx}
\usepackage{natbib}

\newcommand{\hsig}{{\sigma}}
\newcommand{\hsigdag}{{\sigma}^\dagger}

\newcommand{\id}{\mathds{1}}

\newcommand{\mcL}{\mathcal{L}}
\newcommand{\mcK}{\mathcal{K}}

\newcommand{\beq}{\begin{eqnarray}}
\newcommand{\eeq}{\end{eqnarray}}

\begin{document}
\title{Environmental non-additivity and Franck-Condon physics in non-equilibrium quantum systems} 
\date{\today}
\author{Henry Maguire}
\affiliation{{Photon Science Institute \& School of Physics and Astronomy, The University of Manchester, Oxford Road, Manchester M13 9PL, United Kingdom}}
\author{Jake Iles-Smith}
\affiliation{{Photon Science Institute \& School of Physics and Astronomy, The University of Manchester, Oxford Road, Manchester M13 9PL, United Kingdom}}
\affiliation{Department of Physics and Astronomy, University of Sheffield, Hounsfield Road, Sheffield, S3 7RH, United Kingdom}
\author{Ahsan Nazir}
\affiliation{{Photon Science Institute \& School of Physics and Astronomy, The University of Manchester, Oxford Road, Manchester M13 9PL, United Kingdom}}

\begin{abstract}
We show that for a quantum system coupled to both vibrational and electromagnetic environments, enforcing additivity of their combined influences results in non-equilibrium dynamics that does not respect the Franck-Condon principle. We overcome this shortcoming by employing a collective coordinate representation of the vibrational environment, which permits the derivation of a non-additive master equation. When applied to a two-level emitter our treatment predicts decreasing photon emission rates with increasing vibrational coupling, consistent with Franck-Condon physics. In contrast, the additive approximation predicts the emission rate to be completely insensitive to vibrations. We find that non-additivity also plays a key role in the stationary non-equilibrium model behaviour, enabling two-level population inversion under incoherent electromagnetic excitation. 
\end{abstract}

\maketitle

\begin{figure}[t]
	\center
	\includegraphics[width=\columnwidth]{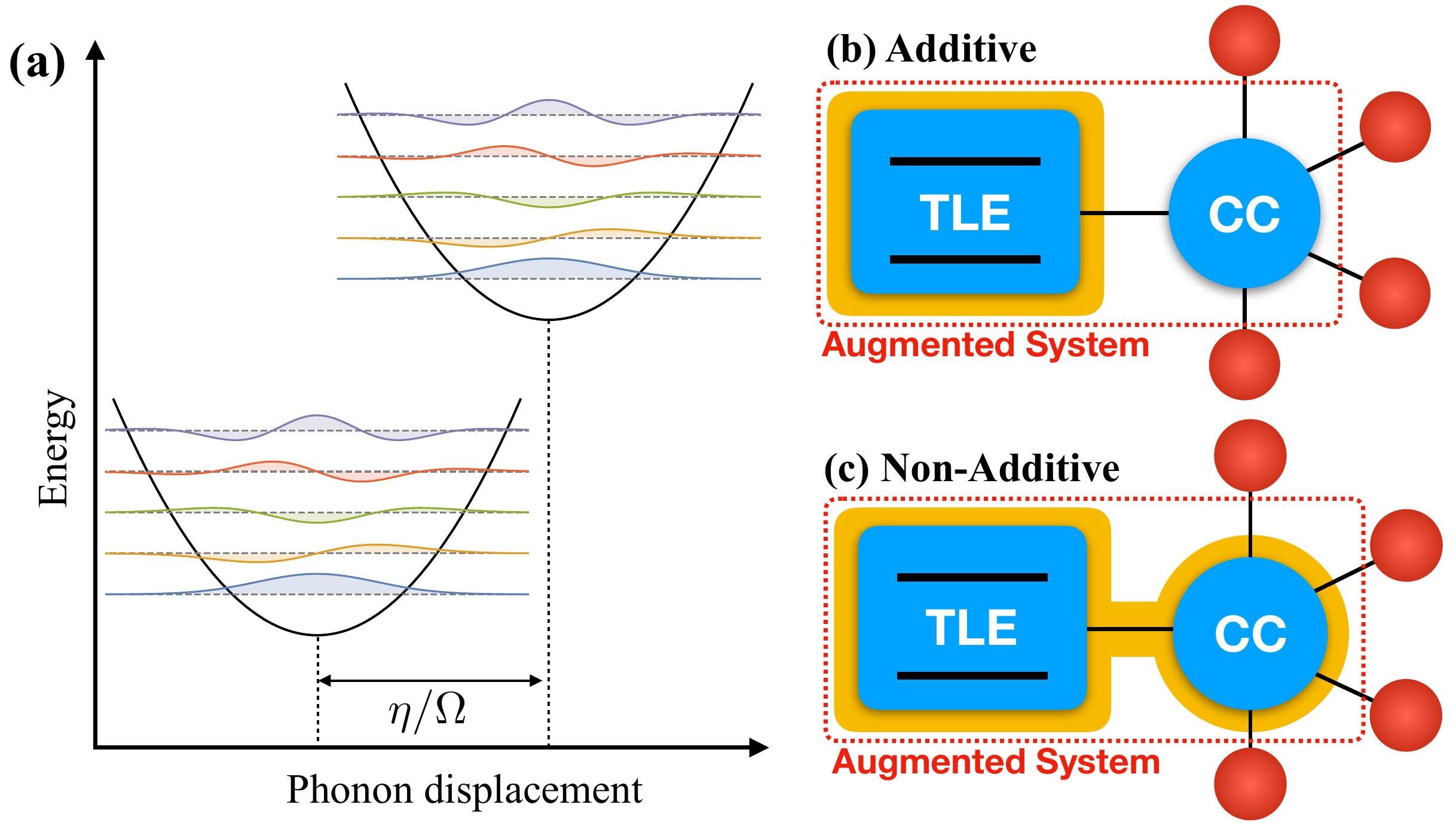}
	\caption{{\bf(a)} Illustration of the vibronic energy structure commonly associated to the Franck-Condon principle. Vibrational coupling leads to the formation of manifolds corresponding to the ground and excited electronic configurations, with 
		transition probabilities  
		proportional to the overlap of the displaced and undisplaced vibrational states.
		{\bf(b), (c)} Schematics of the collective coordinate (CC) mapping. 
		In the additive case {\bf(b)} the electromagnetic field (shaded) is sensitive only to the two-level emitter (TLE), whereas in the non-additive case {\bf(c)} it is sensitive to the full augmented system (TLE+CC).  
		}
	\label{fig:ill}
\end{figure}

The Franck-Condon (FC) principle~\cite{May_and_Kuhn,nitzanbook} is an 
invaluable tool in the study of solid-state and molecular emitters. 
The principle states that electronic transitions of an emitter 
occur without changes to the motions of its nuclei or those of its environment. 
As a result, transition rates become dependent on the overlap between vibrational configurations in the initial and final states, which are generally displaced from one another [see Fig.~\ref{fig:ill}(a)]. 
This picture provides 
an intuitive starting point for studying the complex interactions between the electronic and vibrational degrees of freedom 
of an emitter and its 
environment, for example through rate equations derived from Fermi's Golden Rule~\cite{May_and_Kuhn,nitzanbook}. 


Faithfully representing 
the full  
non-equilibrium dynamics of such systems 
requires moving beyond 
rate equations and 
instead employing an explicitly time-dependent approach. This should be 
non-perturbative in the electron-vibrational coupling and thus capable of 
capturing the dynamical 
influence of vibrational displacement 
on the electronic states. 
Examples 
include polaron~\cite{doi:10.1002/wcms.1111,polaronreview} and collective coordinate~\cite{iles2014environmental,iles2016energy,doi:10.1063/1.5040898} master equations, hierarchical equations of motion~\cite{Tanimura89,Ishizaki05,Ishizaki09}, path integrals~\cite{Makri1, Makri2,PhysRevE.84.041926}, and tensor network methods~\cite{Rosenbach16,schroeder16,Strathearn17}. 
Nevertheless, 
it is interactions with
the electromagnetic environment that 
ultimately give rise to the observed electronic (e.g.~optical) transitions. 
Our focus is thus on the important question of 
how to incorporate 
electromagnetic interactions 
into the  
dynamical formalism,    
such that they respect the non-perturbative nature of the vibrational coupling. 

Given that 
interactions with the electromagnetic field in free space 
are weak, 
it is often assumed 
that the Markovian dynamics 
they generate 
can  be added 
to the equations of motion unmodified due to the presence of vibrations~\cite{doi:10.1021/ct200126d,1367-2630-14-7-073027,doi:10.1021/jz3010317,PhysRevB.86.085302,Ulhaq:13,PhysRevLett.110.217401,PhysRevA.90.063818,doi:10.1063/1.4932307,PhysRevE.94.052101,Barth2016nonHamiltonian,PhysRevA.96.012125,C7SC02983G,doi:10.1063/1.5009114,0953-4075-51-5-054002}.
Though justifiable in certain
circumstances~\cite{PhysRevA.75.013811,1751-8121-40-48-015,PhysRevLett.110.217401,Giusteri2017,PhysRevA.97.062124}, 
additivity is 
in general a stringent requirement~\cite{Giusteri2017,PhysRevA.97.062124,Mitchison2018,1367-2630-19-12-123037,doi:10.1142/S1230161217400108} 
that can break down even if all environments are weakly coupled to the system~\cite{Mitchison2018}. In fact,
we shall show below that 
the dynamics obtained in this manner can exhibit fundamental flaws, 
such as 
disregarding 
the FC principle. 
In certain cases, both the vibrational and electromagnetic environments may be treated non-perturbatively~\cite{Dijkstra14arxiv,delpino18}, 
but this comes at an inevitable cost in terms of computational effort and complexity within the formalism.

Here we seek to retain both the simplicity of the Markovian description of the electromagnetic interactions and a non-perturbative treatment of the electron-vibrational coupling, but without the undesirable additivity restriction. 
This is made possible through a collective coordinate (CC) transformation~\cite{Garg1985,Thoss2001,iles2014environmental,doi:10.1063/1.5040898}, which incorporates 
non-perturbative effects of the vibrational environment into an enlarged (augmented) system 
[see Fig.~\ref{fig:ill}(b),(c)]. 
This in turn enables 
a Markovian master equation to be derived in the eigenbasis of the augmented
system space, rather than that of the original bare emitter, by tracing out the electromagnetic environment 
and residual vibrational modes~\cite{doi:10.1063/1.5040898}. 
On doing so we find that electromagnetic transitions become sensitive to the non-perturbative vibrational dynamics captured by the CC mapping, 
and our procedure thus retains the non-additive effects 
crucial to obtaining quantum dynamics that are consistent with the FC principle [Fig.~\ref{fig:ill}(c)]. 
If instead we enforce additivity [Fig.~\ref{fig:ill}(b)],
the resulting electronic 
decay dynamics becomes independent of the electron-vibrational coupling. We show that  
capturing non-additivity 
is also vital for accurately representing the stationary non-equilibrium behaviour within our model. Specifically, under incoherent electromagnetic excitation the non-additive interplay between the electromagnetic field and vibrations directly enables electronic population inversion for situations impossible within the additive approach.

Before
examining non-equilibrium dynamics 
explicitly, we can illustrate the shortcomings of an additive approximation through arguments based on a simple Fermi Golden Rule calculation. 
We consider a two-level molecular emitter (a monomer) with  
electronic excited state $\ket{e}$ and ground state $\ket{g}$, separated by an energy $\epsilon$ ($\hbar=1$ throughout). 
Coupling to the electromagnetic field induces transitions between the electronic states, which are also assumed to 
couple with strength $\eta$ 
to a single (harmonic) vibrational mode of frequency $\Omega$, leading 
to the formation of a displaced manifold associated to the 
excited electronic configuration. This is the situation depicted qualitatively in Fig.~\ref{fig:ill}(a), though our considerations here and throughout the rest of the paper also apply in the case of continuum phonon environments, where the discrete mode would be identified as the CC post mapping (see below). 

We assume for the purpose of calculating the rate that shortly after excitation the system  
has relaxed to thermal equilibrium in the excited state manifold, $\rho_{\rm eq} = \sum_m p_m\ket{e,\tilde{m}}\!\bra{e,\tilde{m}}$, where $p_m = e^{-m\Omega/k_{\rm B}T}/\sum_n e^{-n\Omega/k_{\rm B}T}$ with 
temperature $T$, and the 
displaced vibrational basis is denoted $\ket{\tilde{m}}=D(\eta/\Omega)\ket{m}$ 
for vibrational Fock state $\ket{m}$ and displacement operator $D(\alpha)$. 
From Fermi's Golden Rule the electronic excited to ground state decay rate is then~\cite{May_and_Kuhn} 
\begin{equation}
\Gamma_{e\rightarrow{}g} = \sum_{n,m} p_n\mathcal{J}(\Delta\omega_{\tilde{m},n})\left\vert\bra{\tilde{m}}\! n \rangle\right\vert^2. 
\end{equation}
There are two principal components to this expression. 
One is the overlap between vibrational configurations, 
$\vert\!\bra{\tilde{m}} n \rangle\!\vert^2$, which is known as the FC factor. 
The other is the electromagnetic spectral density $\mathcal{J}(\omega)$. This describes the system-field coupling strength
weighted by the electromagnetic density of states, and should be sampled at all energy differences between 
relevant states in the excited and ground manifolds, $\Delta\omega_{\tilde{m},n}$. 
In the additive approximation, however, the electromagnetic field coupling is treated in isolation from the vibrational interactions, 
and the electromagnetic spectral density is then incorrectly sampled only at the single frequency $\epsilon$ corresponding to the bare electronic ground and excited state splitting. 
The expression for the emission rate then reduces to $\Gamma_{e\rightarrow{}g} \approx \mathcal{J}_0 \sum_n\left\vert\bra{\tilde{0}} n \rangle\right\vert^2 = \mathcal{J}_0$, where $\mathcal{J}_0=\mathcal{J}(\epsilon)$ and we have used $\sum_n \ket{n}\!\bra{n} = \id$. 
Thus, 
in the additive case the FC factor vanishes, and the transition rate 
loses its dependence on the electron-vibrational coupling.
Note that this reasoning can be used to show that the flat spectral density approximation commonly used in quantum optics theory~\cite{carmichael2009statistical} also fails in regimes of strong coupling to vibrational modes.

We now develop a 
microscopic 
description in order to establish 
the extent to which non-additivity can influence 
the quantum dynamics of electron-vibrational models beyond the heuristic arguments outlined above. Our Hamiltonian is written as $H = H_{\rm S} + H_{\rm I} + H_{\rm B}$, 
with system Hamiltonian 
$H_{\rm S} = \epsilon\ket{e}\!\bra{e}$. The electronic configuration of the emitter molecule is directly influenced by both vibrational and electromagnetic environments, such that  $H_{\rm I} = H_{\rm I}^{\rm PH}  + H_{\rm I}^{\rm EM}$. 
Within the harmonic approximation 
the electron-vibrational coupling is written 
\begin{equation}
H_{\rm I}^{\rm PH} = \ket{e}\!\bra{e}\otimes\sum_k g_k ({b}_{k}^\dagger + {b}_{k})+ \ket{e}\!\bra{e}\sum\limits_k \frac{g_k^2}{\nu_k},
\end{equation}
where ${b}_k$ is the annihilation operator for the $k^{\rm th}$ phonon mode 
and the second term shifts the excited state due to the reorganisation energy associated to vibrational displacement. 
The coupling to the phonon environment is characterised by its spectral density, for which we take the common form $J(\nu) = \sum_k \vert{g_k}\vert^2\delta(\nu-\nu_k) = \alpha\nu_0^2 \gamma \nu/\left[(\nu^2-\nu_0^2)^2 + \gamma^2\nu^2\right]$. 
Here $\alpha$ and $\nu_0$ define the coupling strength and peak position, respectively, and $\gamma$ controls whether $J(\nu)$ is narrow (underdamped) or broad (overdamped)~\cite{Dijkstra15,iles2016energy}. 
In addition to the phonon environment, we also have an explicit coupling to the  electromagnetic field, given by $H_{\rm I}^{\rm EM}=-{\bf d}\cdot{\bf E}$ in the dipole approximation, where ${\bf d}$ is the emitter dipole operator and ${\bf E}$ is the electric field operator~\cite{loudon2000quantum,breuer2002theory,agarwalQO}. Ignoring polarisation degrees of freedom and working in the 
rotating wave approximation, 
this then takes the form
\begin{equation}\label{eq:EM}
H_{\rm I}^{\rm EM} = \sum\limits_l (f_l\hsigdag {a}_l  + f_l^\ast \hsig{a}^\dagger_l),
\end{equation}
where $\hsig=\ket{g}\!\bra{e} $ 
and ${a}_l$ is the annihilation operator for the $l^{\rm th}$ mode of the electromagnetic field. 
The spectral density for the light-matter coupling is defined as 
$\mathcal{J}(\omega) = \sum_l \vert f_l\vert^2 \delta(\omega-\omega_l) = (2\pi \epsilon^3)^{-1}\Gamma_0 \omega^3$~\cite{loudon2000quantum,breuer2002theory,agarwalQO}, where $\Gamma_0$ is the spontaneous emission rate for the two-level emitter
in the absence of phonons.
Finally, $H_{\rm B} = H_{\rm B}^{\rm EM}+H_{\rm B}^{\rm PH}=\sum_l\omega_l {a}_l^\dagger {a}_l + \sum_k \nu_k {b}_k^\dagger {b}_k$ is the sum of the internal Hamiltonians for the electromagnetic and vibrational environments.

Applying a CC mapping to the phonon bath allows us to incorporate its influence on the electronic dynamics non-perturbatively, and in particular to capture the resulting dynamical generation of electron-vibrational correlations~\cite{Garg1985,Thoss2001,iles2014environmental,iles2016energy}.
Our Hamiltonian maps as $H=H_{\rm S} + H_{\rm I} + H_{\rm B}\rightarrow H_{\rm S}^\prime + H_{\rm I}^R +H^{\rm EM}_{\rm I}+ H^R_{\rm B}+H_{\rm B}^{\rm EM}$, which leaves the light-matter coupling unchanged. 
Here, we have introduced 
the transformed Hamiltonians
\begin{align}
H_{\rm S}^\prime &= H_{\rm S} + \eta\ket{e}\!\bra{e}({b}^\dagger + {b}+\pi\alpha/2\eta) + \Omega{b}^\dagger {b},\label{augmentedsys}\\
H_{\rm I}^R &= ({b}^\dagger + {b})\sum_m h_m({c}_m^\dagger + {c}_m) + ({b}^\dagger + {b})^2\sum_m \frac{h_m^2}{\tilde{\nu}_m},\label{residualinteraction}\\
H_{\rm B}^R& = \sum\limits_m\tilde{\nu}_m{c}_m^\dagger {c}_m,
\end{align}
where ${b}+b^{\dagger}=\sum_kg_k({b}_k^{\dagger}+{b}_k)/\eta$ defines creation and annihilation operators for the CC, ${c}_m$ is the annihilation operator for the $m^{\rm th}$ mode of the residual environment to which it couples, and we have expressed the reorganisation energy as $\sum_k{g_k^2}/{\nu_k}=\int_0^{\infty}d\nu J(\nu)/\nu =\pi\alpha/2$. The CC parameters can be written in terms of the quantities defining the vibrational 
spectral density: $\eta^2 =\pi\alpha\nu_0/2$ and $\Omega = \nu_0$~\cite{iles2016energy}. 
Coupling between the augmented emitter-CC system and the residual phonon environment is described by an Ohmic spectral density $J_{\rm R}(\nu) = \sum_m\vert h_m\vert^2\delta(\nu-\tilde{\nu}_m)  =\gamma \nu/2\pi\nu_0$~\cite{iles2016energy}, and ensures that the vibrational environment still acts as a continuum of modes after the mapping. 
As in the single mode case discussed earlier, the coupling to the CC leads to the formation of two vibronic manifolds associated to the ground and excited electronic configurations. 
The coupling to the residual environment induces transitions \emph{within} each vibronic manifold. This leads both to broadening and to dynamical 
relaxation of the phonon environment, which typically occurs on a sub-picosecond timescale.

From the mapped Hamiltonian we 
derive a second-order Born-Markov master equation by tracing over the residual environment and the electromagnetic field~\cite{breuer2002theory}, 
both of which are assumed to remain in thermal equilibrium at temperatures $T_{\rm R}$ and $T_{\rm EM}$, respectively. The resulting master equation can be written 
$\partial_t\rho(t) =\mathcal{L}[\rho(t)]$ with 
Liouvillian~\footnote{For further details see the Supplemental Material}: 
\begin{equation}
\label{eq:liouvillian}
\mathcal{L}[\rho(t)]= -i\left[H^\prime_{\rm S},\rho(t)\right] + \mathcal{K}_{\rm R}[\rho(t)] + \mathcal{K}_{\rm EM}[\rho(t)],
\end{equation}
where $\rho(t)$ is the reduced state of the augmented emitter-CC system. Here, $\mathcal{K}_{\rm R}$ is 
a superoperator representing the action of the residual phonon environment~\cite{iles2016energy}:
\begin{equation}
\mathcal{K}_{\rm R}[\rho(t)] = \left[S, \rho(t)\zeta\right] + \left[\zeta^{\dagger}\rho(t), S\right],
\end{equation}
with $S = {b}^{\dagger}+{b}$ and 
\begin{align}
\label{eq:NewRCOperatorsDiscrete}
\zeta = \frac{\pi}{2}\sum_{jk}J_{\rm R}(\lambda_{jk}) \left[\coth(\frac{ \lambda_{jk}}{2k_{\rm B}T_{\rm R}})+1\right]S_{jk}\dyad{\psi_j}{\psi_k},
\end{align}
where the eigenbasis of the augmented system is defined through $H_{\rm S}^{\prime}\ket{\psi_j} = \psi_j\ket{\psi_j}$, giving $\lambda_{jk} = \psi_j-\psi_k$ and $S_{jk}=\bra{\psi_j}S\ket{\psi_k}$.
We solve for the eigenvalues $\psi_j$ and eigenstates $\ket{\psi_j}$ numerically, taking the basis $\{\ket{g},\ket{e}\}$ for the TLE and a Fock (number) state basis for the CC.

The effects of the electromagnetic field interaction are contained within $\mathcal{K}_{\rm EM}$. 
Importantly, the augmented emitter-CC system Hamiltonian, $H_{\rm S}'$, is treated (numerically) exactly within the formalism.  
This is 
crucial in capturing non-additive effects of the electromagnetic and vibrational environments, 
as it means that when we move the electromagnetic interaction Hamiltonian [Eq.~(\ref{eq:EM})] into the interaction picture, we do so with respect to the full augmented system Hamiltonian $H_{\rm S}^\prime$ [Eq.~(\ref{augmentedsys})]. The mapping thus ensures that the electromagnetic environment is sensitive to the underlying eigenstructure of both the electronic and vibrational states. The superoperator representing the dynamical influence of the electromagnetic environment 
then takes the form: 
\begin{align}
\mcK_{\rm EM}[\rho(t)] &= -\left[\hsigdag,{\chi}_1\rho(t)\right] -\left[\hsig, {\chi}_2\rho(t)\right] + \text{h.c.},
\end{align}  
where we have introduced the rate operators 
$
\chi_1 = \sum_{jk}\sigma_{jk} \Gamma_\downarrow(\lambda_{jk})\ket{\psi_j}\!\bra{\psi_k}
$ 
and 
$
\chi_2 = \sum_{jk}\sigma^\ast_{jk} \Gamma_\uparrow(\lambda_{jk})\ket{\psi_k}\!\bra{\psi_j}
$, 
with transition rates given by $\Gamma_{\downarrow}(\lambda) = \pi\mathcal{J}(\lambda)(n(\lambda) + 1)$ and
$\Gamma_{\uparrow} (\lambda) = \pi\mathcal{J}(\lambda)n(\lambda)$, 
for field occupation number $n(\lambda) = (\exp{\lambda/k_{\rm B}T_{\rm EM}}-1)^{-1}$ and $\sigma_{jk}=\bra{\psi_j}\sigma\ket{\psi_k}$. 
From these expressions, it is evident that the interaction between the system and the electromagnetic field is dependent on the eigenstructure of the augmented system, and thus on the emitter-vibrational coupling through the identification of 
the CC, and its coupling to the electronic system. 
We therefore refer to this theory as being \emph{non-additive}.

In contrast, within the 
additive approach 
$\mcK_{\rm EM}$ is derived without reference to the vibrational coupling. In the present setting, this amounts to neglecting the modification of the system eigenstructure due to vibrational interactions encoded in the mapped system Hamiltonian $H_{\rm S}^\prime$, and instead moving the electromagnetic interaction Hamiltonian [Eq.~(\ref{eq:EM})] into the interaction picture with respect to the original system Hamiltonian $H_{\rm S}$. 
This results in the standard Lindblad dissipator common in quantum optics theory: 
$\mcK_{\rm EM}[\rho(t)] = \frac{\Gamma_0}{2}(n(\epsilon)+1)\mcL_{\sigma}[\rho(t)]+ \frac{\Gamma_0}{2}n(\epsilon)\mcL_{\sigma^\dagger}[\rho(t)]$, 
where $\mcL_{O}[\rho] = 2O\rho O^\dagger - \{O^\dagger O,\rho\}$. 
It is thus clear that within the additive approximation  
the electromagnetic field superoperator loses its explicit dependence on the vibrational environment. 
For vanishing electromagnetic interactions ($\Gamma_0\to 0$), the additive and non-additive theories become equivalent and the problem reduces to the independent boson model, for which an exact solution can be obtained. We verify that the CC master equation agrees with this exact solution for a range of parameters in the Supplemental Material.

\begin{figure}\center
	\includegraphics[width=0.49\columnwidth]{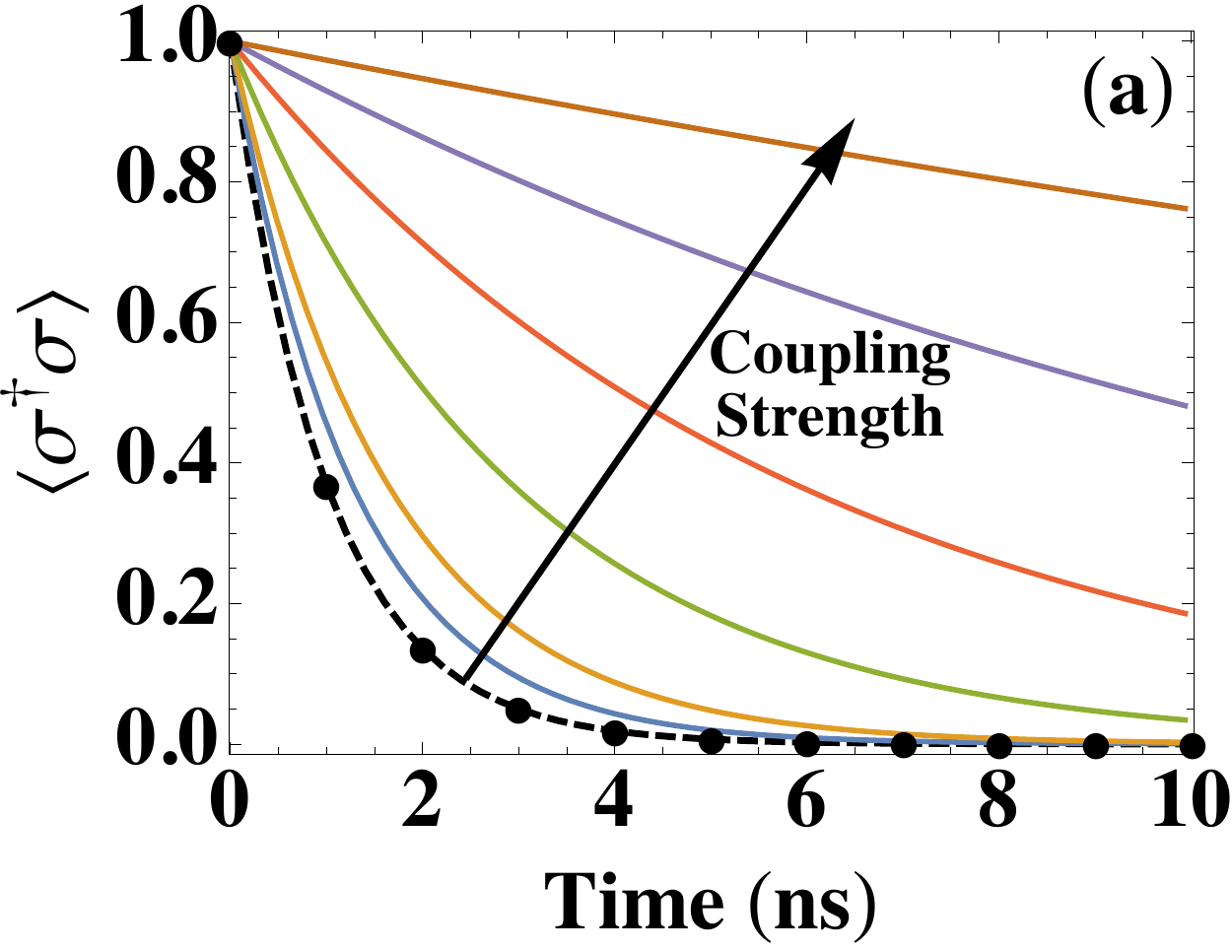}
		\includegraphics[width=0.49\columnwidth]{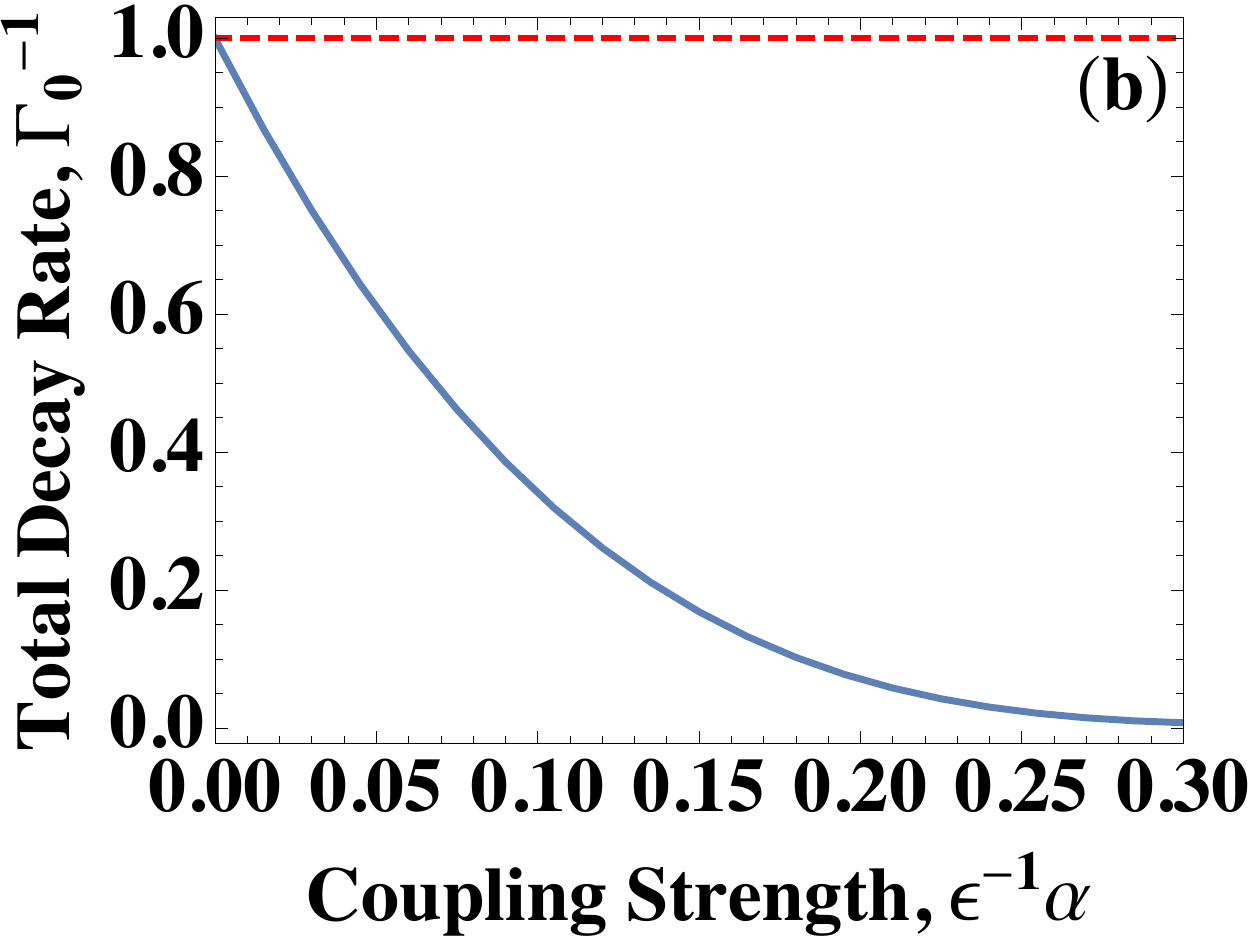}
		\caption{{\bf (a)} Emitter population dynamics from the additive (dots) and non-additive (solid) theories for increasing vibrational coupling strength $\epsilon^{-1}\alpha = 0.025,~0.05,~ 0.1,~ 0.15,~ 0.2, $ and $0.25$. 
		The non-additive theory shows a steady reduction of the decay rate for increasing coupling, whereas within the additive theory the rate remains constant (i.e.~all dotted curves lie on top of each other). 
		{\bf(b)} Excited to ground state emission rate against vibrational coupling strength from the additive (dashed) and non-additive (solid) theories in units of the bare decay rate $\Gamma_0$.
	Parameters: $\epsilon = 8065~$cm$^{-1}$, 
	$\nu_0 = 400~$cm$^{-1}$, $\gamma =80$~cm$^{-1 }$, $\Gamma_0^{-1} =100$~ps, and $T_{\rm R} =T_{\rm EM}= 300$~K.}
\label{fig:dynamics}
\end{figure}

We are now in a position to investigate the 
impact of non-additive effects on the dynamics of our model system. 
We begin by considering the decay of an emitter initialised in its excited state with the collective coordinate in a thermal state set by the residual bath temperature $T_{\rm R}$: $\rho(0) = \ket{e}\!\bra{e}\otimes\rho_{th}$, where $\rho_{th} = \exp(-\Omega {b}^\dagger{b}/k_{\rm B}T_{\rm R})/\tr[\exp(-\Omega {b}^\dagger{b}/k_{\rm B}T_{\rm R})]$. 
This approximates a canonical thermal state of the original vibrational Hamiltonian in the unmapped representation at the same temperature, and 
is thus consistent with rapid (vertical) excitation of the system whereby the electronic state changes suddenly  
but the vibrational states 
remain unchanged~\cite{Note1}. 
The vibrational environment will subsequently relax towards the displaced thermal state associated to the excited state manifold, captured dynamically within our approach.

Fig.~\ref{fig:dynamics}(a) shows the emitter excited state population dynamics predicted by the additive (dotted) and non-additive (solid) theories for increasing electron-phonon coupling at ambient temperature. 
Both theories give rise to exponential decay, with the rate in the additive theory remaining constant across all electron-phonon coupling strengths. The non-additive theory, in contrast, displays a monotonic decrease in the decay rate with increasing phonon coupling. This can be seen explicitly in Fig.~\ref{fig:dynamics}(b), where we extract the decay rates directly from the master equation.  
Specifically, 
the excited to ground state transition rate can be written as
\begin{equation}
\Gamma_{e\rightarrow g} =\sum_n \bra{g, n}\mathcal{L}[\rho_X(0)]\ket{g,n},
\end{equation}
with the Liouvillian taken to be additive or non-additive depending on which case is under investigation. Here we must modify the initial state to account for the aforementioned rapid residual bath induced relaxation of the CC to a displaced thermal state prior to emission: $\rho_X(0) = \ket{e}\!\bra{e}\otimes e^{-X}\rho_{th}e^X$, where $X = \Omega^{-1}\eta(b^\dagger-b)$. 
As expected, the rate from the additive theory displays no variation with phonon coupling strength, 
in line with the simple Golden Rule calculation discussed previously but at odds with the FC principle. This again highlights 
deficiencies with the phenomenological additive treatment of the electromagnetic field. 
Conversely, the non-additive theory shows a steady reduction of the emission rate as a function of phonon coupling, consistent with FC physics. As the displacement between the ground and excited state manifolds increases linearly with the electron-phonon coupling strength, this reduces the overlap between the vibrational states and thus suppresses electromagnetic transitions.

\begin{figure}[t]
	\center
	\includegraphics[width =0.49 \columnwidth]{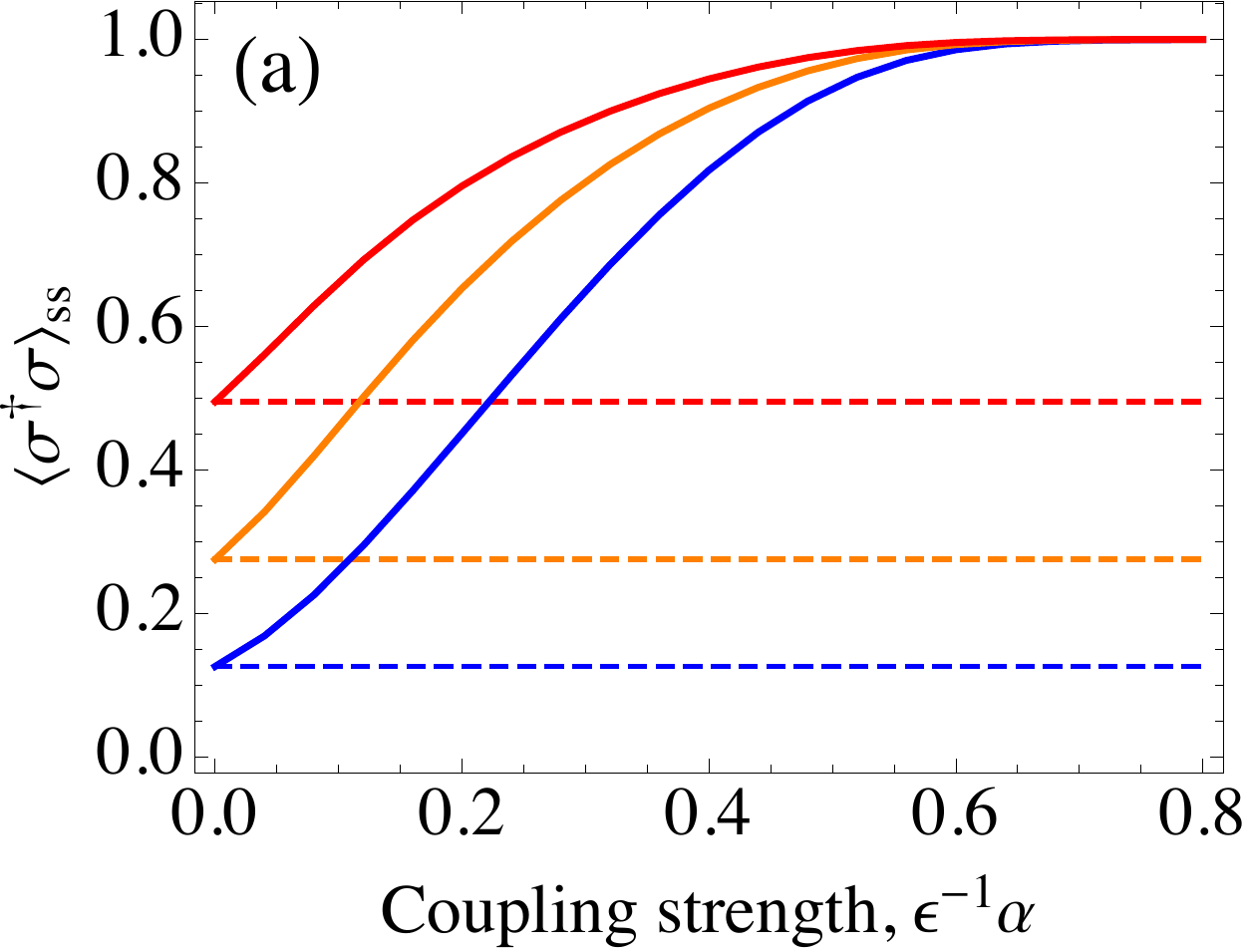}
	\includegraphics[width=  0.49\columnwidth]{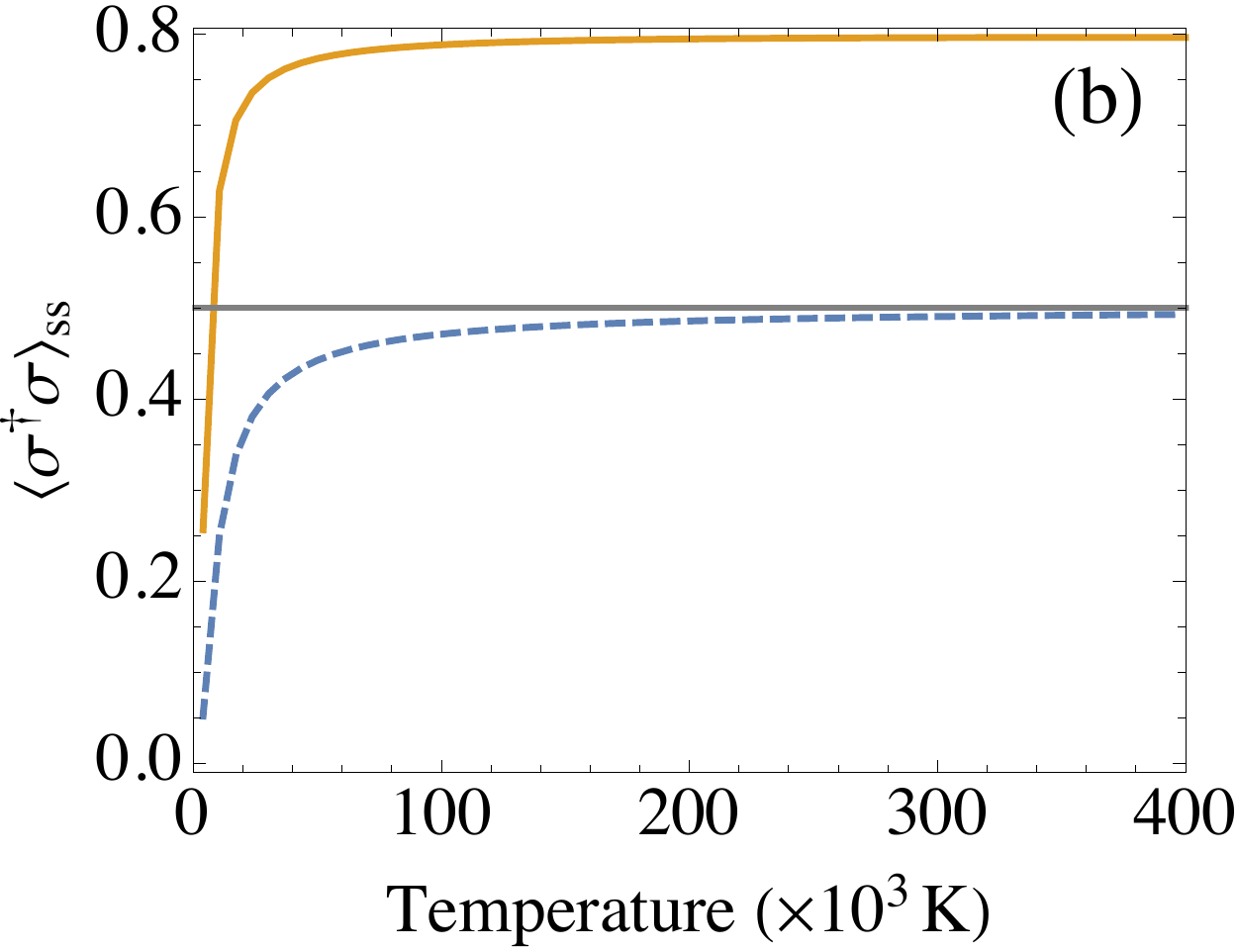}	
	\caption{(a) Steady-state emitter population as a function of the electron-phonon coupling strength for the additive (dashed) and non-additive (solid) theories. The electromagnetic field temperatures are $T_{\rm EM} = 6000$~K (blue, lower), $12000$~K (orange, middle), and $60000$~K (red, upper).
	(b) Steady-state emitter population with varying temperature for $\alpha = 0.3\epsilon$ from the additive (dashed) and non-additive (solid) theories. In the additive theory the stationary population asymptotically approaches 0.5 (grey line) and never displays an inversion, in contrast to the non-additive treatment. Other parameters are as in Fig.~\ref{fig:dynamics}.}
	\label{fig:inc}
\end{figure}

It is important to stress that discrepancies between the additive and non-additive treatments in our model extend further than spontaneous emission processes. For example, we now consider situations in which the emitter is driven incoherently via thermal occupation of the electromagnetic environment, which constitutes an important building block of widely used models for natural and artificial solar energy conversion. 
Fig.~\ref{fig:inc}(a) shows the steady state population of the electronic excited state 
as a function of electron-phonon coupling strength, where 
the additive treatment once again displays no variation, simply matching the equilibrium distribution expected in the absence of vibrations. The non-additive treatment, on the other hand, shows a monotonic increase in the steady state population of the excited state manifold. Most strikingly, at large coupling strengths there emerges a steady state population inversion. 
In the absence of phonons, such an inversion would be impossible, with emission and absorption processes balancing each other in equilibrium. 
This remains true in the presence of phonons when the electromagnetic field is treated additively, as highlighted in Fig.~\ref{fig:inc}(b). Here, the additive theory approaches, but never exceeds, a maximum steady state population $\langle\hsigdag\hsig\rangle = 0.5$ in the limit of very large temperatures. In contrast, within the non-additive theory, cooperative effects between the electromagnetic and vibrational environments lead to non-equilibrium stationary states that display substantial levels of population inversion. That such effects should be possible, even for continuum environments, is made clear from the CC mapping. 
Within the non-additive theory the electromagnetic field has access to the full vibrational structure of the emitter, providing the necessary states to drive a population inversion. 
This points to a crucial difference between non-additive and additive treatments, where disregarding the eigenstructure of the combined electronic and vibrational system misses key aspects of the non-equilibrium physics.

In summary, we have demonstrated that for
models of electronic systems strongly coupled to vibrational environments, including the electromagnetic field 
in an 
additive manner 
can lead to dynamics inconsistent with the FC principle.
By developing a dynamical formalism based on collective coordinate mappings, we capture the impact of non-additive effects to recover both transient and stationary non-equilibrium behaviour consistent with FC physics. 
Furthermore, we find that 
for common model assumptions on the forms of vibrational and electromagnetic couplings, non-additive phenomena enable 
steady-state population inversion under incoherent electromagnetic excitation conditions.   
Our findings open up a number of avenues for further exploration in both natural and artificial light harvesting, for example 
whether non-additive effects may be harnessed to enhance work extraction in models of quantum heat engines and 
solar energy conversion devices.

\emph{Acknowledgements.---}The authors wish to thank Adam Stokes and Neill Lambert for useful discussions. H.M. is supported by the EPSRC. A.N. and J.I.S. are supported by the EPSRC, grant no.~EP/N008154/1. J.I.S. also acknowledges support from the Royal Commission for the Exhibition of 1851.


%

\widetext

\begin{appendix}
\section{Supplemental Material}

\subsection{Further details of the master equation derivation}
\subsubsection{Non-Additive}
In this supplement we outline further details of the master equations used in the main text. We begin by considering the non-additive case, whereby, after the collective-coordinate (CC) mapping, our interaction Hamiltonian is given by the sum of residual phonon bath and electromagnetic field coupling terms [Eqs.~(3) and~(5) in the main text], $H_{\rm I} = S\otimes B+S^2\sum_m{h_m^2}/{\tilde{\nu}_m}+\sum_{\alpha}A_\alpha\otimes E_\alpha$. Here, $S={b}^\dagger+{b}$ and $B=\sum_m h_m({c}_m^\dagger + {c}_m)$ describe the CC-residual bath interaction, while $A_1 = \hsigdag$, $A_2 = \hsig$, $E_1 = \sum_l f_l {a}_l$ and $E_2 = \sum_l f^\ast_l  {a}_l^\dagger$ define the coupling of the two-level emitter (TLE) to the electromagnetic field. We now move into the interaction picture with respect to the augmented system Hamiltonian describing the coupled TLE and CC, $H_{\rm S}'=H_{\rm S} + \eta\ket{e}\!\bra{e}({b}^\dagger + {b}+\pi\alpha/2\eta) + \Omega{b}^\dagger {b}$, plus the residual and electromagnetic bath Hamiltonians $H_{\rm B}^R = \sum_m\tilde{\nu}_m{c}_m^\dagger {c}_m$ and $H_{\rm B}^{\rm EM}=\sum_l\omega_l {a}_l^\dagger {a}_l$, respectively. This gives 
\begin{equation}
H_{\rm I}(t) = S(t)\otimes B(t)+S(t)^2\sum_m\frac{h_m^2}{\tilde{\nu}_m}+\sum_{\alpha}A_\alpha(t)\otimes E_\alpha(t), 
\end{equation}
where $S(t)=e^{iH_{\rm S}'t}Se^{-iH_{\rm S}'t}$, $A_\alpha(t)=e^{iH_{\rm S}'t}A_\alpha e^{-iH_{\rm S}'t}$, $B(t)=\sum_m h_m({c}_m^\dagger e^{i\tilde{\nu}_mt} + {c}_me^{-i\tilde{\nu}_mt})$, $E_1 = \sum_l f_l {a}_le^{-i \omega_l t}$ and $E_2 = \sum_l f^\ast_l  {a}_l^\dagger e^{i \omega_l t}$. Within the interaction picture, we then follow the standard procedure to derive a Redfield master equation, tracing out the residual and electromagnetic environments within the Born-Markov approximations~\cite{breuer2002theory}. Moving back into the Schr\"odinger picture, the resulting master equation may be written in the general form
\begin{equation}
\partial_t\rho(t)=\mathcal{L}[\rho(t)]= -i\left[H^\prime_{\rm S},\rho(t)\right] + \mathcal{K}_{\rm R}[\rho(t)] + \mathcal{K}_{\rm EM}[\rho(t)],
\end{equation}
where $\rho(t)$ is the reduced density operator of the augmented system, from which either the TLE or CC dynamics may be obtained by tracing out the relevant degrees of freedom. The superoperators $\mathcal{K}_{\rm R}$ and $ \mathcal{K}_{\rm EM}$ encode, respectively, the influence of the residual bath and the electromagnetic field interactions on the augmented system dynamics. Note that due to the Born-Markov approximations there are no mixed terms between the residual phonon bath and the electromagnetic field in the master equation above. Nevertheless, our master equation is still non-additive with respect to the \emph{original} phonon environment and the electromagnetic field due to the CC mapping, which incorporates non-perturbative vibrational effects into the enlarged augmented system Hamiltonian $H_{\rm S}'$ used to move into the interaction picture. This results in an electromagnetic superoperator that has explicit dependence on the form and strength of the system-vibrational coupling, as we shall see below. 

Starting with the influence of the residual phonon environment, we find
\begin{equation}\label{residualgeneral}
\mathcal{K}_{\rm R}[\rho(t)] =-i\sum_m\frac{h_m^2}{\tilde{\nu}_m}[S^2,\rho(t)] - \int_{0}^{\infty}d\tau \big[S, \big[S(-\tau),\rho(t)\big]\big]C_+(\tau) -\int_{0}^{\infty}d\tau \big[S, \big\{S(-\tau),\rho(t)\big\}\big]C_- (\tau),
\end{equation}
with residual bath correlation functions defined as
\begin{equation}
C_{\pm}(\tau) = \frac{1}{2}\langle B(\tau)B\pm B(-\tau)B \rangle,
\end{equation}
where the expectation value is taken with respect to a thermal state at temperature $T_{\rm R}$. These can be found given the form of $B(t)$ above to be
\begin{equation}
C_+(\tau) = \int_{0}^{\infty}d\omega J_{\rm R}(\omega)\coth\left(\frac{\omega}{2k_{\rm B}T_{\rm R}}\right)\cos\omega\tau \quad \text{and}  \quad C_-(\tau) = \int_{0}^{\infty}d\omega J_{\rm R}(\omega)\sin\omega\tau.
\end{equation}
Here, we have defined the residual bath spectral density $J_{\rm R}(\nu) = \sum_m\vert h_m\vert^2\delta(\nu-\tilde{\nu}_m)$ as in the main text. 
We now decompose the system operators into the eigenbasis of the augmented TLE-CC Hamiltonian, $H_{\rm S}'\ket{\psi_j} = \psi_j\ket{\psi_j}$, giving $S(t) = \sum_{jk}S_{jk} e^{i\lambda_{jk}t}\dyad{\psi_j}{\psi_k}$, where $S_{jk}=\bra{\psi_j}S\ket{\psi_k}$ and $\lambda_{jk}=\psi_j-\psi_k$. 
Inserting this decomposition into Eq.~(\ref{residualgeneral}) and exchanging the order of integration over time and frequency, we can now perform the integral over $\tau$ using the Sokhotski-Plemelj theorem, $\int_0^{\infty}d\tau e^{\pm i\epsilon \tau}=\pi\delta(\epsilon)\pm iP(1/\epsilon)$, where $P$ stands for the Cauchy principal value.
After some algebra, and neglecting imaginary contributions as justified in Refs.~\cite{iles2014environmental,iles2016energy}, we find that the residual bath superoperator may be written
\begin{equation}
\mathcal{K}_{\rm R}[\rho(t)] = \left[S, \rho(t)\zeta\right] + \left[\zeta^{\dagger}\rho(t), S\right],
\end{equation}
with 
\begin{align}
\label{eq:NewRCOperatorsDiscrete}
\zeta = \frac{\pi}{2}\sum_{jk}J_{\rm R}(\lambda_{jk}) \left[\coth(\frac{ \lambda_{jk}}{2k_{\rm B}T_{\rm R}})+1\right]S_{jk}\dyad{\psi_j}{\psi_k},
\end{align}
as given in the main text.

Next, 
we consider the superoperator for the electromagnetic field, which can be written 
\begin{equation}\label{eq:meq}
\begin{split}
\mathcal{K}_{\rm EM}[\rho(t)] =&-\int\limits_0^\infty d\tau\left(
\left[\hsigdag,\hsig(-\tau)\rho(t)\right]C_{12}(\tau) 
+ \left[\rho(t)\hsig(-\tau),\hsigdag(t)\right]C_{21}(-\tau)\right)
\\
&
-\int\limits_0^\infty d\tau
\left(
\left[\hsig,\hsigdag(-\tau)\rho(t)\right]C_{21}(\tau) 
+ \left[\rho(t)\hsigdag(-\tau),\hsig(t)\right]C_{12}(-\tau)
\right).
\end{split}
\end{equation} 
Here, the bath correlation functions are defined as $C_{\alpha\alpha'}(\tau)=\langle E_\alpha(\tau)E_{\alpha'}\rangle$ with the expectation value taken with respect to a thermal state at temperature $T_{\rm EM}$. This gives
\begin{equation}
C_{12}(\tau) = \int\limits_0^\infty d\tau \mathcal{J}(\omega)(n(\omega) + 1)e^{-i\omega\tau}\quad \text{and} \quad 
C_{21}(\tau) = \int\limits_0^\infty d\tau \mathcal{J}(\omega)n(\omega) e^{i\omega\tau},
\end{equation}
where we have defined the electromagnetic field spectral density $\mathcal{J}(\omega) = \sum_l \vert f_l\vert^2 \delta(\omega-\omega_l)$ and the field occupation number $n(\omega) = (\exp{\omega/k_{\rm B}T_{\rm EM}}-1)^{-1}$, again as in the main text. 

For the non-additive treatment of the electromagnetic field interactions we again decompose the system operators into the eigenbasis of the augmented TLE-CC Hamiltonian 
such that $\sigma(t) = \sum_{jk}\sigma_{jk}e^{i\lambda_{jk}t}\ket{\psi_j}\!\bra{\psi_k}$, where $\sigma_{jk}=\bra{\psi_j}\sigma\ket{\psi_k}$. 
Inserting this decomposition into Eq.~(\ref{eq:meq}) and exchanging the order of the time and frequency integrals, we perform the integral over $\tau$ using the Sokhotski-Plemelj theorem as before. Again, after some algebra and neglecting small imaginary terms, we find
\begin{align}\label{EMsuperoperator}
\mathcal{K}_{\rm EM}[\rho(t)]=-\left[\hsigdag,\chi_1 \rho(t)\right] - \left[\hsig,\chi_2 \rho(t)\right] + \operatorname{h.c.},
\end{align}  
with rate operators 
\begin{align}
\chi_1& = \sum_{jk}\sigma_{jk} \Gamma_\downarrow(\lambda_{jk})\ket{\psi_j}\!\bra{\psi_k},\\
\chi_2& = \sum_{jk}\sigma^\ast_{jk} \Gamma_\uparrow(\lambda_{jk})\ket{\psi_k}\!\bra{\psi_j},
\end{align}
where $\Gamma_{\downarrow}(\lambda) = \pi\mathcal{J}(\lambda)(n(\lambda) + 1)$ and
$\Gamma_{\uparrow} (\lambda) = \pi\mathcal{J}(\lambda)n(\lambda)$ as in the main text. Note that these non-additive rate operators account for the full eigenstructure of the electronic and CC degrees of freedom.

\subsubsection{Additive}

Within the additive master equation the description of the residual phonon bath is unchanged, as $\mathcal{K}_{\rm R}[\rho(t)]$ has no dependence on the electromagnetic field coupling. However, the electromagnetic field superoperator $\mathcal{K}_{\rm EM}[\rho(t)]$ is altered significantly, as it is no longer sensitive to the (non-perturbative) vibrational coupling.

Specifically, to get $\mathcal{K}_{\rm EM}[\rho(t)]$ in the additive case, one should completely ignore the presence of vibrations when moving the relevant system operators $\sigma$ and $\sigma^{\dagger}$ into the interaction picture. This results in $\sigma(t) = e^{iH_{\rm S}t}\sigma e^{-iH_{\rm S}t}=\sigma e^{-i\epsilon t}$ and $\sigma^{\dagger}(t) = e^{iH_{\rm S}t}\sigma^{\dagger} e^{-iH_{\rm S}t}= \sigma^{\dagger} e^{i\epsilon\tau}$, where $H_{\rm S}=\epsilon\ket{e}\bra{e}$. The impact is a simplification of the rate operators to
\begin{align}
\chi_1^{add.}  =  \Gamma_\downarrow(\epsilon)\sigma,
\end{align}
and 
\begin{align}
\chi_2^{add.} = \Gamma_\uparrow(\epsilon)\sigma^{\dagger},
\end{align}
which when inserted into Eq.~(\ref{EMsuperoperator}) 
results in the standard Lindblad form common in quantum optics theory: 
\begin{align}
\mcK_{\rm EM}[\rho(t)] = \frac{\Gamma_0}{2}(n(\epsilon)+1)\mcL_{\sigma}[\rho(t)]+ \frac{\Gamma_0}{2}n(\epsilon)\mcL_{\sigma^\dagger}[\rho(t)],
\end{align} 
with $\mcL_{O}[\rho] = 2O\rho O^\dagger - \{O^\dagger O,\rho\}$.

\begin{figure}[t]\center
	\includegraphics[width=0.49\columnwidth]{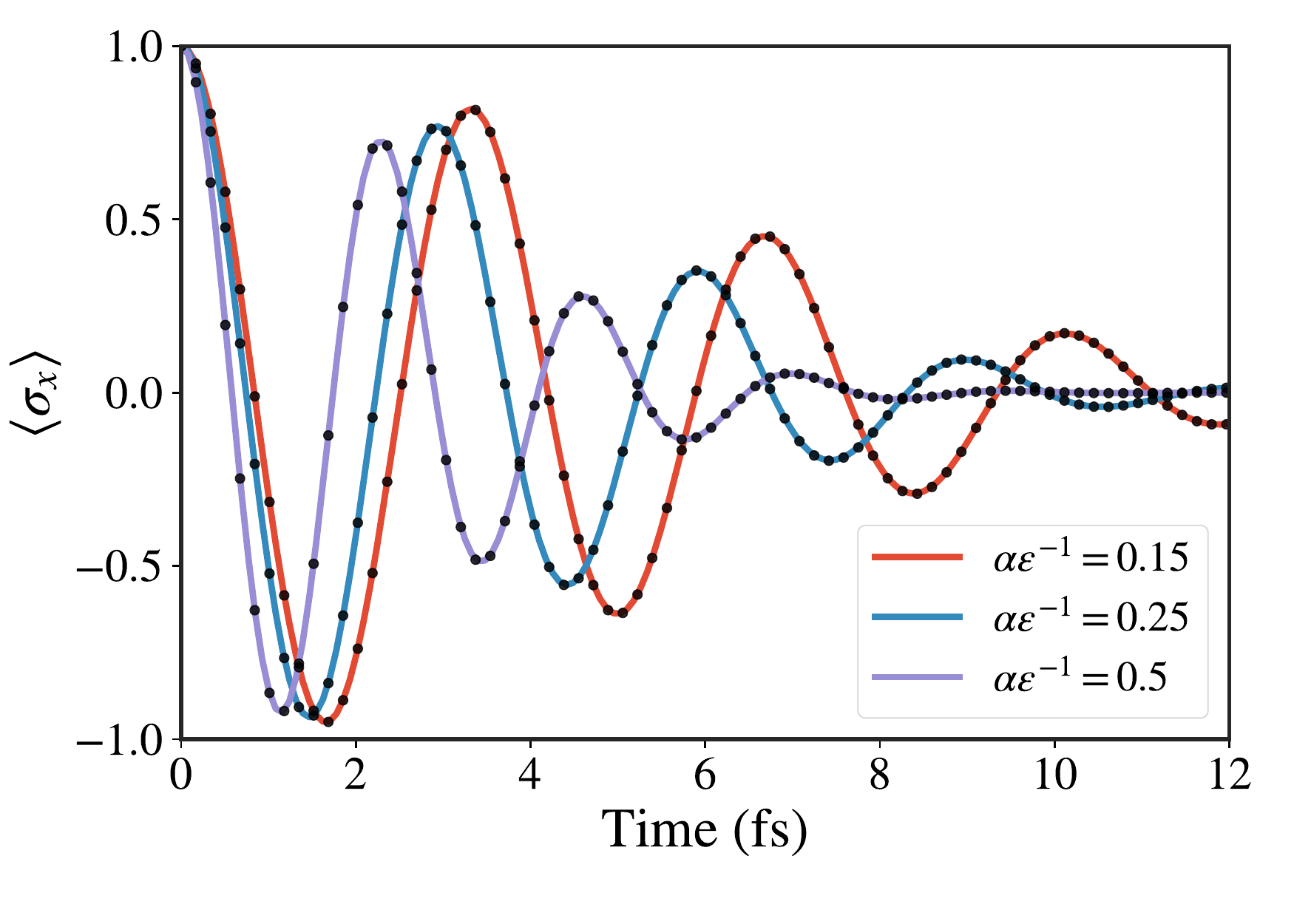}
	\includegraphics[width=0.49\columnwidth]{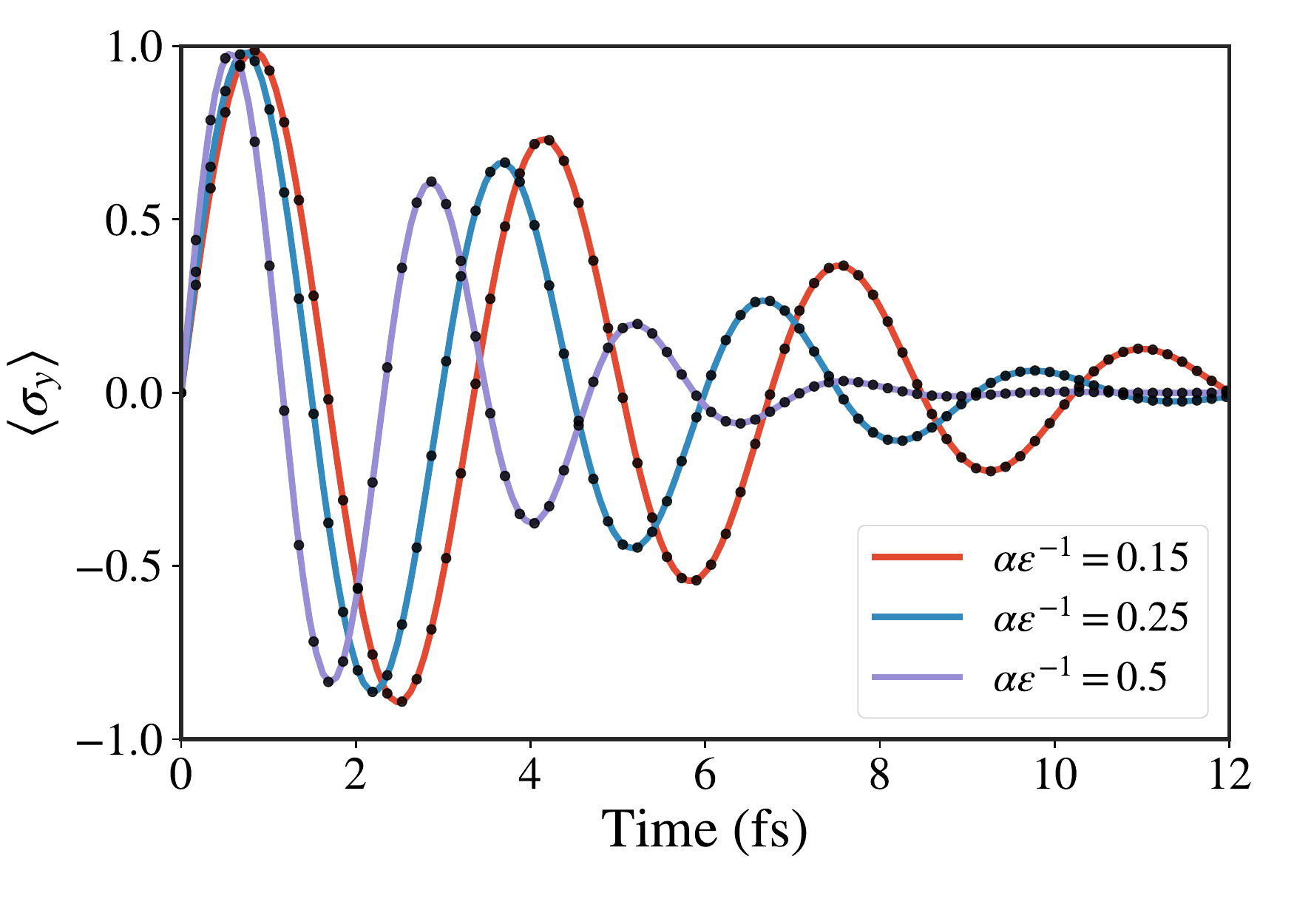}
	\caption{Two-level system coherence dynamics ($\langle \sigma_x\rangle=\rho_{eg}(t)+\rho_{ge}(t)$ and $\langle \sigma_y\rangle=i(\rho_{eg}(t)-\rho_{ge}(t))$) as a function of time for the independent boson model in the strong-coupling regime: exact solution (dots) and collective coordinate master equation (solid curves).
	Parameters: $\epsilon = 8065~$cm$^{-1}$, 
	$\nu_0 = 400~$cm$^{-1}$, $\gamma =80$~cm$^{-1 }$, and $T_{\rm R} = 300$~K.}
	\label{fig:IBMdynamics}
\end{figure}

\subsection{Independent boson model: comparison of the collective coordinate master equation and exact solutions}
In the absence of coupling to an optical environment ($\Gamma_0=0$), the model outlined in the manuscript reduces to the independent boson model 
\begin{equation}\label{HIBM}
H_{\rm }^{\rm IBM} = \left(\epsilon + \sum\limits_k \frac{g_k^2}{\nu_k}\right)\ket{e}\!\bra{e} +  \ket{e}\!\bra{e}\otimes\sum_k g_k ({b}_{k}^\dagger + {b}_{k}) + \sum_k \omega_k {b}_{k}^\dagger{b}_{k},
\end{equation}
which is exactly solvable~\cite{breuer2002theory}, for example via a polaron transformation of the form $H^{\prime} = U^{\dagger}HU$, where $U=\ket{e}\!\bra{e}\prod_{k} D(-g_k/\omega_k)+\ket{g}\!\bra{g}$. This removes the interaction term, 
permitting a partial trace to be taken over the phonon modes, which we consider here to be initialised in the thermal state 
\begin{equation}
\label{eq:environmentState}
\rho_{\rm PH}(0) = \exp(-\sum_k \omega_k {b}_{k}^\dagger{b}_{k}/k_{\rm B}T_R)/\tr[\exp(-\sum_k \omega_k {b}_{k}^\dagger{b}_{k}/k_{\rm B}T_R)], 
\end{equation}
with total initial density matrix $\chi(0)=\rho_{\rm S}(0)\rho_{\rm PH}(0)$. Since the interaction term in Eq.~(\ref{HIBM}) commutes with the system Hamiltonian, within the independent boson model the phonon bath causes no transitions between electronic eigenstates and thus the system populations are static, $\rho_{gg}(t) = \rho_{gg}(0)$ and $\rho_{ee}(t) = \rho_{ee}(0)$, where $\rho_{ij}(t)=\langle i|\rho_{\rm S}(t)|j\rangle$ for $i,j\in\{g,e\}$ are system density matrix elements.
The system coherences do evolve, and are governed by 
\begin{equation}
\rho_{eg}(t) = \rho_{eg}(0)e^{-i \epsilon t}e^{-\Gamma(t)} \quad \textnormal{and} \quad \rho_{ge}(t) = (\rho_{eg}(t))^*,
\end{equation}
where the decoherence function is defined as
\begin{equation}
\Gamma(t) = -\int_{0}^{\infty}d\omega \frac{J(\omega)}{\omega^2} \coth(\frac{\omega}{2k_BT})(1-\cos(\omega t)).
\end{equation}
The dynamics of these exact solutions can be compared to the collective coordinate master equation used in the main manuscript by simply setting the electromagnetic coupling $\Gamma_0$ to zero in the latter. In Fig.~\ref{fig:IBMdynamics} above we compare the two approaches using the same parameters as in Fig.~2 of the manuscript (though with $\Gamma_0=0$) and find that the collective coordinate master equation matches the exact solution well into the strong phonon-coupling regime. The initial condition for the two-level system is taken to be $\rho_{\rm S}(0) = \frac{1}{2}(\ket{e}+\ket{g})\!(\bra{e}+\bra{g})$. For the exact solution, the environment is initialised in the thermal state $\rho_{\rm PH}(0)$ of Eq.~(\ref{eq:environmentState}), as stated before, while for the collective coordinate master equation the CC is initialised in the thermal state $\rho_{th} = \exp(-\Omega {b}^\dagger{b}/k_{\rm B}T_{\rm R})/\tr[\exp(-\Omega {b}^\dagger{b}/k_{\rm B}T_{\rm R})]$ with the residual environment held in thermal equilibrium at temperature $T_R$ throughout the dynamics.


\begin{thebibliography}{47}%
\makeatletter
\providecommand \@ifxundefined [1]{%
 \@ifx{#1\undefined}
}%
\providecommand \@ifnum [1]{%
 \ifnum #1\expandafter \@firstoftwo
 \else \expandafter \@secondoftwo
 \fi
}%
\providecommand \@ifx [1]{%
 \ifx #1\expandafter \@firstoftwo
 \else \expandafter \@secondoftwo
 \fi
}%
\providecommand \natexlab [1]{#1}%
\providecommand \enquote  [1]{``#1''}%
\providecommand \bibnamefont  [1]{#1}%
\providecommand \bibfnamefont [1]{#1}%
\providecommand \citenamefont [1]{#1}%
\providecommand \href@noop [0]{\@secondoftwo}%
\providecommand \href [0]{\begingroup \@sanitize@url \@href}%
\providecommand \@href[1]{\@@startlink{#1}\@@href}%
\providecommand \@@href[1]{\endgroup#1\@@endlink}%
\providecommand \@sanitize@url [0]{\catcode `\\12\catcode `\$12\catcode
  `\&12\catcode `\#12\catcode `\^12\catcode `\_12\catcode `\%12\relax}%
\providecommand \@@startlink[1]{}%
\providecommand \@@endlink[0]{}%
\providecommand \url  [0]{\begingroup\@sanitize@url \@url }%
\providecommand \@url [1]{\endgroup\@href {#1}{\urlprefix }}%
\providecommand \urlprefix  [0]{URL }%
\providecommand \Eprint [0]{\href }%
\providecommand \doibase [0]{http://dx.doi.org/}%
\providecommand \selectlanguage [0]{\@gobble}%
\providecommand \bibinfo  [0]{\@secondoftwo}%
\providecommand \bibfield  [0]{\@secondoftwo}%
\providecommand \translation [1]{[#1]}%
\providecommand \BibitemOpen [0]{}%
\providecommand \bibitemStop [0]{}%
\providecommand \bibitemNoStop [0]{.\EOS\space}%
\providecommand \EOS [0]{\spacefactor3000\relax}%
\providecommand \BibitemShut  [1]{\csname bibitem#1\endcsname}%
\let\auto@bib@innerbib\@empty
\bibitem [{\citenamefont {{May}}\ and\ \citenamefont
  {{K{\"u}hn}}(2004)}]{May_and_Kuhn}%
  \BibitemOpen
  \bibfield  {author} {\bibinfo {author} {\bibfnamefont {V.}~\bibnamefont
  {{May}}}\ and\ \bibinfo {author} {\bibfnamefont {O.}~\bibnamefont
  {{K{\"u}hn}}},\ }\href@noop {} {\emph {\bibinfo {title} {Charge and Energy
  Transfer Dynamics in Molecular Systems}}}\ (\bibinfo  {publisher} {Wiley},\
  \bibinfo {year} {2004})\BibitemShut {NoStop}%
\bibitem [{\citenamefont {Nitzan}(2006)}]{nitzanbook}%
  \BibitemOpen
  \bibfield  {author} {\bibinfo {author} {\bibfnamefont {A.}~\bibnamefont
  {Nitzan}},\ }\href@noop {} {\emph {\bibinfo {title} {Chemical Dynamics in
  Condensed Phases}}}\ (\bibinfo  {publisher} {Oxford University Press},\
  \bibinfo {year} {2006})\BibitemShut {NoStop}%
\bibitem [{\citenamefont {Jang}\ and\ \citenamefont
  {Cheng}(2012)}]{doi:10.1002/wcms.1111}%
  \BibitemOpen
  \bibfield  {author} {\bibinfo {author} {\bibfnamefont {S.}~\bibnamefont
  {Jang}}\ and\ \bibinfo {author} {\bibfnamefont {Y.-C.}\ \bibnamefont
  {Cheng}},\ }\href@noop {} {\bibfield  {journal} {\bibinfo  {journal} {WIREs
  Comput. Mol. Sci.}\ }\textbf {\bibinfo {volume} {3}},\ \bibinfo {pages} {84}
  (\bibinfo {year} {2012})}\BibitemShut {NoStop}%
\bibitem [{\citenamefont {Nazir}\ and\ \citenamefont
  {McCutcheon}(2016)}]{polaronreview}%
  \BibitemOpen
  \bibfield  {author} {\bibinfo {author} {\bibfnamefont {A.}~\bibnamefont
  {Nazir}}\ and\ \bibinfo {author} {\bibfnamefont {D.~P.~S.}\ \bibnamefont
  {McCutcheon}},\ }\href@noop {} {\bibfield  {journal} {\bibinfo  {journal} {J.
  Phys.: Condens. Matter}\ }\textbf {\bibinfo {volume} {28}},\ \bibinfo {pages}
  {103002} (\bibinfo {year} {2016})}\BibitemShut {NoStop}%
\bibitem [{\citenamefont {Iles-Smith}\ \emph {et~al.}(2014)\citenamefont
  {Iles-Smith}, \citenamefont {Lambert},\ and\ \citenamefont
  {Nazir}}]{iles2014environmental}%
  \BibitemOpen
  \bibfield  {author} {\bibinfo {author} {\bibfnamefont {J.}~\bibnamefont
  {Iles-Smith}}, \bibinfo {author} {\bibfnamefont {N.}~\bibnamefont {Lambert}},
  \ and\ \bibinfo {author} {\bibfnamefont {A.}~\bibnamefont {Nazir}},\
  }\href@noop {} {\bibfield  {journal} {\bibinfo  {journal} {Phys. Rev. A}\
  }\textbf {\bibinfo {volume} {90}},\ \bibinfo {pages} {032114} (\bibinfo
  {year} {2014})}\BibitemShut {NoStop}%
\bibitem [{\citenamefont {Iles-Smith}\ \emph {et~al.}(2016)\citenamefont
  {Iles-Smith}, \citenamefont {Dijkstra}, \citenamefont {Lambert},\ and\
  \citenamefont {Nazir}}]{iles2016energy}%
  \BibitemOpen
  \bibfield  {author} {\bibinfo {author} {\bibfnamefont {J.}~\bibnamefont
  {Iles-Smith}}, \bibinfo {author} {\bibfnamefont {A.~G.}\ \bibnamefont
  {Dijkstra}}, \bibinfo {author} {\bibfnamefont {N.}~\bibnamefont {Lambert}}, \
  and\ \bibinfo {author} {\bibfnamefont {A.}~\bibnamefont {Nazir}},\
  }\href@noop {} {\bibfield  {journal} {\bibinfo  {journal} {J. Chem. Phys.}\
  }\textbf {\bibinfo {volume} {144}},\ \bibinfo {pages} {044110} (\bibinfo
  {year} {2016})}\BibitemShut {NoStop}%
\bibitem [{\citenamefont {Wertnik}\ \emph {et~al.}(2018)\citenamefont
  {Wertnik}, \citenamefont {Chin}, \citenamefont {Nori},\ and\ \citenamefont
  {Lambert}}]{doi:10.1063/1.5040898}%
  \BibitemOpen
  \bibfield  {author} {\bibinfo {author} {\bibfnamefont {M.}~\bibnamefont
  {Wertnik}}, \bibinfo {author} {\bibfnamefont {A.}~\bibnamefont {Chin}},
  \bibinfo {author} {\bibfnamefont {F.}~\bibnamefont {Nori}}, \ and\ \bibinfo
  {author} {\bibfnamefont {N.}~\bibnamefont {Lambert}},\ }\href@noop {}
  {\bibfield  {journal} {\bibinfo  {journal} {J. Chem. Phys.}\ }\textbf
  {\bibinfo {volume} {149}},\ \bibinfo {pages} {084112} (\bibinfo {year}
  {2018})}\BibitemShut {NoStop}%
\bibitem [{\citenamefont {Tanimura}\ and\ \citenamefont
  {Kubo}(1989)}]{Tanimura89}%
  \BibitemOpen
  \bibfield  {author} {\bibinfo {author} {\bibfnamefont {Y.}~\bibnamefont
  {Tanimura}}\ and\ \bibinfo {author} {\bibfnamefont {R.}~\bibnamefont
  {Kubo}},\ }\href {\doibase 10.1143/JPSJ.58.101} {\bibfield  {journal}
  {\bibinfo  {journal} {J. Phys. Soc. Jpn.}\ }\textbf {\bibinfo {volume}
  {58}},\ \bibinfo {pages} {101} (\bibinfo {year} {1989})}\BibitemShut
  {NoStop}%
\bibitem [{\citenamefont {Ishizaki}\ and\ \citenamefont
  {Tanimura}(2005)}]{Ishizaki05}%
  \BibitemOpen
  \bibfield  {author} {\bibinfo {author} {\bibfnamefont {A.}~\bibnamefont
  {Ishizaki}}\ and\ \bibinfo {author} {\bibfnamefont {Y.}~\bibnamefont
  {Tanimura}},\ }\href {\doibase 10.1143/JPSJ.74.3131} {\bibfield  {journal}
  {\bibinfo  {journal} {J. Phys. Soc. Jpn.}\ }\textbf {\bibinfo {volume}
  {74}},\ \bibinfo {pages} {3131} (\bibinfo {year} {2005})}\BibitemShut
  {NoStop}%
\bibitem [{\citenamefont {Ishizaki}\ and\ \citenamefont
  {Fleming}(2009)}]{Ishizaki09}%
  \BibitemOpen
  \bibfield  {author} {\bibinfo {author} {\bibfnamefont {A.}~\bibnamefont
  {Ishizaki}}\ and\ \bibinfo {author} {\bibfnamefont {G.~R.}\ \bibnamefont
  {Fleming}},\ }\href {\doibase 10.1063/1.3155372} {\bibfield  {journal}
  {\bibinfo  {journal} {J. Chem. Phys.}\ }\textbf {\bibinfo {volume} {130}},\
  \bibinfo {pages} {234111} (\bibinfo {year} {2009})}\BibitemShut {NoStop}%
\bibitem [{\citenamefont {Makri}\ and\ \citenamefont
  {Makarov}(1995{\natexlab{a}})}]{Makri1}%
  \BibitemOpen
  \bibfield  {author} {\bibinfo {author} {\bibfnamefont {N.}~\bibnamefont
  {Makri}}\ and\ \bibinfo {author} {\bibfnamefont {D.~E.}\ \bibnamefont
  {Makarov}},\ }\href {\doibase 10.1063/1.469508} {\bibfield  {journal}
  {\bibinfo  {journal} {J. Chem. Phys.}\ }\textbf {\bibinfo {volume} {102}},\
  \bibinfo {pages} {4600} (\bibinfo {year} {1995}{\natexlab{a}})}\BibitemShut
  {NoStop}%
\bibitem [{\citenamefont {Makri}\ and\ \citenamefont
  {Makarov}(1995{\natexlab{b}})}]{Makri2}%
  \BibitemOpen
  \bibfield  {author} {\bibinfo {author} {\bibfnamefont {N.}~\bibnamefont
  {Makri}}\ and\ \bibinfo {author} {\bibfnamefont {D.~E.}\ \bibnamefont
  {Makarov}},\ }\href {\doibase 10.1063/1.469509} {\bibfield  {journal}
  {\bibinfo  {journal} {J. Chem. Phys.}\ }\textbf {\bibinfo {volume} {102}},\
  \bibinfo {pages} {4611} (\bibinfo {year} {1995}{\natexlab{b}})}\BibitemShut
  {NoStop}%
\bibitem [{\citenamefont {Nalbach}\ \emph {et~al.}(2011)\citenamefont
  {Nalbach}, \citenamefont {Braun},\ and\ \citenamefont
  {Thorwart}}]{PhysRevE.84.041926}%
  \BibitemOpen
  \bibfield  {author} {\bibinfo {author} {\bibfnamefont {P.}~\bibnamefont
  {Nalbach}}, \bibinfo {author} {\bibfnamefont {D.}~\bibnamefont {Braun}}, \
  and\ \bibinfo {author} {\bibfnamefont {M.}~\bibnamefont {Thorwart}},\ }\href
  {\doibase 10.1103/PhysRevE.84.041926} {\bibfield  {journal} {\bibinfo
  {journal} {Phys. Rev. E}\ }\textbf {\bibinfo {volume} {84}},\ \bibinfo
  {pages} {041926} (\bibinfo {year} {2011})}\BibitemShut {NoStop}%
\bibitem [{\citenamefont {Rosenbach}\ \emph {et~al.}(2016)\citenamefont
  {Rosenbach}, \citenamefont {Cerrillo}, \citenamefont {Huelga}, \citenamefont
  {Cao},\ and\ \citenamefont {Plenio}}]{Rosenbach16}%
  \BibitemOpen
  \bibfield  {author} {\bibinfo {author} {\bibfnamefont {R.}~\bibnamefont
  {Rosenbach}}, \bibinfo {author} {\bibfnamefont {J.}~\bibnamefont {Cerrillo}},
  \bibinfo {author} {\bibfnamefont {S.~F.}\ \bibnamefont {Huelga}}, \bibinfo
  {author} {\bibfnamefont {J.}~\bibnamefont {Cao}}, \ and\ \bibinfo {author}
  {\bibfnamefont {M.~B.}\ \bibnamefont {Plenio}},\ }\href
  {http://stacks.iop.org/1367-2630/18/i=2/a=023035} {\bibfield  {journal}
  {\bibinfo  {journal} {New J. Phys.}\ }\textbf {\bibinfo {volume} {18}},\
  \bibinfo {pages} {023035} (\bibinfo {year} {2016})}\BibitemShut {NoStop}%
\bibitem [{\citenamefont {Schr\"{o}der}\ and\ \citenamefont
  {Chin}(2016)}]{schroeder16}%
  \BibitemOpen
  \bibfield  {author} {\bibinfo {author} {\bibfnamefont {F.~A. Y.~N.}\
  \bibnamefont {Schr\"{o}der}}\ and\ \bibinfo {author} {\bibfnamefont {A.~W.}\
  \bibnamefont {Chin}},\ }\href@noop {} {\bibfield  {journal} {\bibinfo
  {journal} {Phys. Rev. B}\ }\textbf {\bibinfo {volume} {93}},\ \bibinfo
  {pages} {075105} (\bibinfo {year} {2016})}\BibitemShut {NoStop}%
\bibitem [{\citenamefont {{Strathearn}}\ \emph {et~al.}(2018)\citenamefont
  {{Strathearn}}, \citenamefont {{Kirton}}, \citenamefont {{Kilda}},
  \citenamefont {{Keeling}},\ and\ \citenamefont {{Lovett}}}]{Strathearn17}%
  \BibitemOpen
  \bibfield  {author} {\bibinfo {author} {\bibfnamefont {A.}~\bibnamefont
  {{Strathearn}}}, \bibinfo {author} {\bibfnamefont {P.}~\bibnamefont
  {{Kirton}}}, \bibinfo {author} {\bibfnamefont {D.}~\bibnamefont {{Kilda}}},
  \bibinfo {author} {\bibfnamefont {J.}~\bibnamefont {{Keeling}}}, \ and\
  \bibinfo {author} {\bibfnamefont {B.~W.}\ \bibnamefont {{Lovett}}},\
  }\href@noop {} {\bibfield  {journal} {\bibinfo  {journal} {Nature Comms.}\
  }\textbf {\bibinfo {volume} {9}},\ \bibinfo {pages} {3322} (\bibinfo {year}
  {2018})}\BibitemShut {NoStop}%
\bibitem [{\citenamefont {Kreisbeck}\ \emph {et~al.}(2011)\citenamefont
  {Kreisbeck}, \citenamefont {Kramer}, \citenamefont {Rodr\'iguez},\ and\
  \citenamefont {Hein}}]{doi:10.1021/ct200126d}%
  \BibitemOpen
  \bibfield  {author} {\bibinfo {author} {\bibfnamefont {C.}~\bibnamefont
  {Kreisbeck}}, \bibinfo {author} {\bibfnamefont {T.}~\bibnamefont {Kramer}},
  \bibinfo {author} {\bibfnamefont {M.}~\bibnamefont {Rodr\'iguez}}, \ and\
  \bibinfo {author} {\bibfnamefont {B.}~\bibnamefont {Hein}},\ }\href@noop {}
  {\bibfield  {journal} {\bibinfo  {journal} {J. Chem. Theory Comput.}\
  }\textbf {\bibinfo {volume} {7}},\ \bibinfo {pages} {2166} (\bibinfo {year}
  {2011})}\BibitemShut {NoStop}%
\bibitem [{\citenamefont {Dijkstra}\ and\ \citenamefont
  {Tanimura}(2012)}]{1367-2630-14-7-073027}%
  \BibitemOpen
  \bibfield  {author} {\bibinfo {author} {\bibfnamefont {A.~G.}\ \bibnamefont
  {Dijkstra}}\ and\ \bibinfo {author} {\bibfnamefont {Y.}~\bibnamefont
  {Tanimura}},\ }\href {http://stacks.iop.org/1367-2630/14/i=7/a=073027}
  {\bibfield  {journal} {\bibinfo  {journal} {New J. Phys.}\ }\textbf {\bibinfo
  {volume} {14}},\ \bibinfo {pages} {073027} (\bibinfo {year}
  {2012})}\BibitemShut {NoStop}%
\bibitem [{\citenamefont {Fassioli}\ \emph {et~al.}(2012)\citenamefont
  {Fassioli}, \citenamefont {Olaya-Castro},\ and\ \citenamefont
  {Scholes}}]{doi:10.1021/jz3010317}%
  \BibitemOpen
  \bibfield  {author} {\bibinfo {author} {\bibfnamefont {F.}~\bibnamefont
  {Fassioli}}, \bibinfo {author} {\bibfnamefont {A.}~\bibnamefont
  {Olaya-Castro}}, \ and\ \bibinfo {author} {\bibfnamefont {G.~D.}\
  \bibnamefont {Scholes}},\ }\href {\doibase 10.1021/jz3010317} {\bibfield
  {journal} {\bibinfo  {journal} {J. Phys. Chem. Lett.}\ }\textbf {\bibinfo
  {volume} {3}},\ \bibinfo {pages} {3136} (\bibinfo {year} {2012})}\BibitemShut
  {NoStop}%
\bibitem [{\citenamefont {Kaer}\ \emph {et~al.}(2012)\citenamefont {Kaer},
  \citenamefont {Nielsen}, \citenamefont {Lodahl}, \citenamefont {Jauho},\ and\
  \citenamefont {M\o{}rk}}]{PhysRevB.86.085302}%
  \BibitemOpen
  \bibfield  {author} {\bibinfo {author} {\bibfnamefont {P.}~\bibnamefont
  {Kaer}}, \bibinfo {author} {\bibfnamefont {T.~R.}\ \bibnamefont {Nielsen}},
  \bibinfo {author} {\bibfnamefont {P.}~\bibnamefont {Lodahl}}, \bibinfo
  {author} {\bibfnamefont {A.-P.}\ \bibnamefont {Jauho}}, \ and\ \bibinfo
  {author} {\bibfnamefont {J.}~\bibnamefont {M\o{}rk}},\ }\href@noop {}
  {\bibfield  {journal} {\bibinfo  {journal} {Phys. Rev. B}\ }\textbf {\bibinfo
  {volume} {86}},\ \bibinfo {pages} {085302} (\bibinfo {year}
  {2012})}\BibitemShut {NoStop}%
\bibitem [{\citenamefont {Ulhaq}\ \emph {et~al.}(2013)\citenamefont {Ulhaq},
  \citenamefont {Weiler}, \citenamefont {Roy}, \citenamefont {Ulrich},
  \citenamefont {Jetter}, \citenamefont {Hughes},\ and\ \citenamefont
  {Michler}}]{Ulhaq:13}%
  \BibitemOpen
  \bibfield  {author} {\bibinfo {author} {\bibfnamefont {A.}~\bibnamefont
  {Ulhaq}}, \bibinfo {author} {\bibfnamefont {S.}~\bibnamefont {Weiler}},
  \bibinfo {author} {\bibfnamefont {C.}~\bibnamefont {Roy}}, \bibinfo {author}
  {\bibfnamefont {S.~M.}\ \bibnamefont {Ulrich}}, \bibinfo {author}
  {\bibfnamefont {M.}~\bibnamefont {Jetter}}, \bibinfo {author} {\bibfnamefont
  {S.}~\bibnamefont {Hughes}}, \ and\ \bibinfo {author} {\bibfnamefont
  {P.}~\bibnamefont {Michler}},\ }\href@noop {} {\bibfield  {journal} {\bibinfo
   {journal} {Opt. Express}\ }\textbf {\bibinfo {volume} {21}},\ \bibinfo
  {pages} {4382} (\bibinfo {year} {2013})}\BibitemShut {NoStop}%
\bibitem [{\citenamefont {McCutcheon}\ and\ \citenamefont
  {Nazir}(2013)}]{PhysRevLett.110.217401}%
  \BibitemOpen
  \bibfield  {author} {\bibinfo {author} {\bibfnamefont {D.~P.~S.}\
  \bibnamefont {McCutcheon}}\ and\ \bibinfo {author} {\bibfnamefont
  {A.}~\bibnamefont {Nazir}},\ }\href {\doibase 10.1103/PhysRevLett.110.217401}
  {\bibfield  {journal} {\bibinfo  {journal} {Phys. Rev. Lett.}\ }\textbf
  {\bibinfo {volume} {110}},\ \bibinfo {pages} {217401} (\bibinfo {year}
  {2013})}\BibitemShut {NoStop}%
\bibitem [{\citenamefont {Betzholz}\ \emph {et~al.}(2014)\citenamefont
  {Betzholz}, \citenamefont {Torres},\ and\ \citenamefont
  {Bienert}}]{PhysRevA.90.063818}%
  \BibitemOpen
  \bibfield  {author} {\bibinfo {author} {\bibfnamefont {R.}~\bibnamefont
  {Betzholz}}, \bibinfo {author} {\bibfnamefont {J.~M.}\ \bibnamefont
  {Torres}}, \ and\ \bibinfo {author} {\bibfnamefont {M.}~\bibnamefont
  {Bienert}},\ }\href@noop {} {\bibfield  {journal} {\bibinfo  {journal} {Phys.
  Rev. A}\ }\textbf {\bibinfo {volume} {90}},\ \bibinfo {pages} {063818}
  (\bibinfo {year} {2014})}\BibitemShut {NoStop}%
\bibitem [{\citenamefont {Killoran}\ \emph {et~al.}(2015)\citenamefont
  {Killoran}, \citenamefont {Huelga},\ and\ \citenamefont
  {Plenio}}]{doi:10.1063/1.4932307}%
  \BibitemOpen
  \bibfield  {author} {\bibinfo {author} {\bibfnamefont {N.}~\bibnamefont
  {Killoran}}, \bibinfo {author} {\bibfnamefont {S.~F.}\ \bibnamefont
  {Huelga}}, \ and\ \bibinfo {author} {\bibfnamefont {M.~B.}\ \bibnamefont
  {Plenio}},\ }\href@noop {} {\bibfield  {journal} {\bibinfo  {journal} {J.
  Chem. Phys.}\ }\textbf {\bibinfo {volume} {143}},\ \bibinfo {pages} {155102}
  (\bibinfo {year} {2015})}\BibitemShut {NoStop}%
\bibitem [{\citenamefont {Chen}\ \emph {et~al.}(2016)\citenamefont {Chen},
  \citenamefont {Chiu},\ and\ \citenamefont {Chen}}]{PhysRevE.94.052101}%
  \BibitemOpen
  \bibfield  {author} {\bibinfo {author} {\bibfnamefont {H.-B.}\ \bibnamefont
  {Chen}}, \bibinfo {author} {\bibfnamefont {P.-Y.}\ \bibnamefont {Chiu}}, \
  and\ \bibinfo {author} {\bibfnamefont {Y.-N.}\ \bibnamefont {Chen}},\ }\href
  {\doibase 10.1103/PhysRevE.94.052101} {\bibfield  {journal} {\bibinfo
  {journal} {Phys. Rev. E}\ }\textbf {\bibinfo {volume} {94}},\ \bibinfo
  {pages} {052101} (\bibinfo {year} {2016})}\BibitemShut {NoStop}%
\bibitem [{\citenamefont {Barth}\ \emph {et~al.}(2016)\citenamefont {Barth},
  \citenamefont {Vagov},\ and\ \citenamefont {Axt}}]{Barth2016nonHamiltonian}%
  \BibitemOpen
  \bibfield  {author} {\bibinfo {author} {\bibfnamefont {A.~M.}\ \bibnamefont
  {Barth}}, \bibinfo {author} {\bibfnamefont {A.}~\bibnamefont {Vagov}}, \ and\
  \bibinfo {author} {\bibfnamefont {V.~M.}\ \bibnamefont {Axt}},\ }\href
  {\doibase 10.1103/PhysRevB.94.125439} {\bibfield  {journal} {\bibinfo
  {journal} {Phys. Rev. B}\ }\textbf {\bibinfo {volume} {94}},\ \bibinfo
  {pages} {125439} (\bibinfo {year} {2016})}\BibitemShut {NoStop}%
\bibitem [{\citenamefont {Qin}\ \emph {et~al.}(2017)\citenamefont {Qin},
  \citenamefont {Shen}, \citenamefont {Zhao},\ and\ \citenamefont
  {Yi}}]{PhysRevA.96.012125}%
  \BibitemOpen
  \bibfield  {author} {\bibinfo {author} {\bibfnamefont {M.}~\bibnamefont
  {Qin}}, \bibinfo {author} {\bibfnamefont {H.~Z.}\ \bibnamefont {Shen}},
  \bibinfo {author} {\bibfnamefont {X.~L.}\ \bibnamefont {Zhao}}, \ and\
  \bibinfo {author} {\bibfnamefont {X.~X.}\ \bibnamefont {Yi}},\ }\href
  {\doibase 10.1103/PhysRevA.96.012125} {\bibfield  {journal} {\bibinfo
  {journal} {Phys. Rev. A}\ }\textbf {\bibinfo {volume} {96}},\ \bibinfo
  {pages} {012125} (\bibinfo {year} {2017})}\BibitemShut {NoStop}%
\bibitem [{\citenamefont {Stones}\ \emph {et~al.}(2017)\citenamefont {Stones},
  \citenamefont {Hossein-Nejad}, \citenamefont {van Grondelle},\ and\
  \citenamefont {Olaya-Castro}}]{C7SC02983G}%
  \BibitemOpen
  \bibfield  {author} {\bibinfo {author} {\bibfnamefont {R.}~\bibnamefont
  {Stones}}, \bibinfo {author} {\bibfnamefont {H.}~\bibnamefont
  {Hossein-Nejad}}, \bibinfo {author} {\bibfnamefont {R.}~\bibnamefont {van
  Grondelle}}, \ and\ \bibinfo {author} {\bibfnamefont {A.}~\bibnamefont
  {Olaya-Castro}},\ }\href {\doibase 10.1039/C7SC02983G} {\bibfield  {journal}
  {\bibinfo  {journal} {Chem. Sci.}\ }\textbf {\bibinfo {volume} {8}},\
  \bibinfo {pages} {6871} (\bibinfo {year} {2017})}\BibitemShut {NoStop}%
\bibitem [{\citenamefont {Juh\'asz}\ and\ \citenamefont
  {Csurgay}(2018)}]{doi:10.1063/1.5009114}%
  \BibitemOpen
  \bibfield  {author} {\bibinfo {author} {\bibfnamefont {I.~B.}\ \bibnamefont
  {Juh\'asz}}\ and\ \bibinfo {author} {\bibfnamefont {A.~I.}\ \bibnamefont
  {Csurgay}},\ }\href@noop {} {\bibfield  {journal} {\bibinfo  {journal} {AIP
  Advances}\ }\textbf {\bibinfo {volume} {8}},\ \bibinfo {pages} {045318}
  (\bibinfo {year} {2018})}\BibitemShut {NoStop}%
\bibitem [{\citenamefont {Chan}\ \emph {et~al.}(2018)\citenamefont {Chan},
  \citenamefont {Gamel}, \citenamefont {Fleming},\ and\ \citenamefont
  {Whaley}}]{0953-4075-51-5-054002}%
  \BibitemOpen
  \bibfield  {author} {\bibinfo {author} {\bibfnamefont {H.~C.~H.}\
  \bibnamefont {Chan}}, \bibinfo {author} {\bibfnamefont {O.~E.}\ \bibnamefont
  {Gamel}}, \bibinfo {author} {\bibfnamefont {G.~R.}\ \bibnamefont {Fleming}},
  \ and\ \bibinfo {author} {\bibfnamefont {K.~B.}\ \bibnamefont {Whaley}},\
  }\href {http://stacks.iop.org/0953-4075/51/i=5/a=054002} {\bibfield
  {journal} {\bibinfo  {journal} {J. Phys. B: At. Mol. Opt. Phys}\ }\textbf
  {\bibinfo {volume} {51}},\ \bibinfo {pages} {054002} (\bibinfo {year}
  {2018})}\BibitemShut {NoStop}%
\bibitem [{\citenamefont {Scala}\ \emph
  {et~al.}(2007{\natexlab{a}})\citenamefont {Scala}, \citenamefont {Militello},
  \citenamefont {Messina}, \citenamefont {Piilo},\ and\ \citenamefont
  {Maniscalco}}]{PhysRevA.75.013811}%
  \BibitemOpen
  \bibfield  {author} {\bibinfo {author} {\bibfnamefont {M.}~\bibnamefont
  {Scala}}, \bibinfo {author} {\bibfnamefont {B.}~\bibnamefont {Militello}},
  \bibinfo {author} {\bibfnamefont {A.}~\bibnamefont {Messina}}, \bibinfo
  {author} {\bibfnamefont {J.}~\bibnamefont {Piilo}}, \ and\ \bibinfo {author}
  {\bibfnamefont {S.}~\bibnamefont {Maniscalco}},\ }\href {\doibase
  10.1103/PhysRevA.75.013811} {\bibfield  {journal} {\bibinfo  {journal} {Phys.
  Rev. A}\ }\textbf {\bibinfo {volume} {75}},\ \bibinfo {pages} {013811}
  (\bibinfo {year} {2007}{\natexlab{a}})}\BibitemShut {NoStop}%
\bibitem [{\citenamefont {Scala}\ \emph
  {et~al.}(2007{\natexlab{b}})\citenamefont {Scala}, \citenamefont {Militello},
  \citenamefont {Messina}, \citenamefont {Maniscalco}, \citenamefont {Piilo},\
  and\ \citenamefont {Suominen}}]{1751-8121-40-48-015}%
  \BibitemOpen
  \bibfield  {author} {\bibinfo {author} {\bibfnamefont {M.}~\bibnamefont
  {Scala}}, \bibinfo {author} {\bibfnamefont {B.}~\bibnamefont {Militello}},
  \bibinfo {author} {\bibfnamefont {A.}~\bibnamefont {Messina}}, \bibinfo
  {author} {\bibfnamefont {S.}~\bibnamefont {Maniscalco}}, \bibinfo {author}
  {\bibfnamefont {J.}~\bibnamefont {Piilo}}, \ and\ \bibinfo {author}
  {\bibfnamefont {K.-A.}\ \bibnamefont {Suominen}},\ }\href
  {http://stacks.iop.org/1751-8121/40/i=48/a=015} {\bibfield  {journal}
  {\bibinfo  {journal} {J. Phys. A: Math. Theor.}\ }\textbf {\bibinfo {volume}
  {40}},\ \bibinfo {pages} {14527} (\bibinfo {year}
  {2007}{\natexlab{b}})}\BibitemShut {NoStop}%
\bibitem [{\citenamefont {Giusteri}\ \emph {et~al.}(2017)\citenamefont
  {Giusteri}, \citenamefont {Recrosi}, \citenamefont {Schaller},\ and\
  \citenamefont {Celardo}}]{Giusteri2017}%
  \BibitemOpen
  \bibfield  {author} {\bibinfo {author} {\bibfnamefont {G.~G.}\ \bibnamefont
  {Giusteri}}, \bibinfo {author} {\bibfnamefont {F.}~\bibnamefont {Recrosi}},
  \bibinfo {author} {\bibfnamefont {G.}~\bibnamefont {Schaller}}, \ and\
  \bibinfo {author} {\bibfnamefont {G.~L.}\ \bibnamefont {Celardo}},\ }\href
  {\doibase 10.1103/PhysRevE.96.012113} {\bibfield  {journal} {\bibinfo
  {journal} {Phys. Rev. E}\ }\textbf {\bibinfo {volume} {96}},\ \bibinfo
  {pages} {012113} (\bibinfo {year} {2017})}\BibitemShut {NoStop}%
\bibitem [{\citenamefont {Ko\l{}ody\ifmmode~\acute{n}\else \'{n}\fi{}ski}\
  \emph {et~al.}(2018)\citenamefont {Ko\l{}ody\ifmmode~\acute{n}\else
  \'{n}\fi{}ski}, \citenamefont {Brask}, \citenamefont {Perarnau-Llobet},\ and\
  \citenamefont {Bylicka}}]{PhysRevA.97.062124}%
  \BibitemOpen
  \bibfield  {author} {\bibinfo {author} {\bibfnamefont {J.}~\bibnamefont
  {Ko\l{}ody\ifmmode~\acute{n}\else \'{n}\fi{}ski}}, \bibinfo {author}
  {\bibfnamefont {J.~B.}\ \bibnamefont {Brask}}, \bibinfo {author}
  {\bibfnamefont {M.}~\bibnamefont {Perarnau-Llobet}}, \ and\ \bibinfo {author}
  {\bibfnamefont {B.}~\bibnamefont {Bylicka}},\ }\href@noop {} {\bibfield
  {journal} {\bibinfo  {journal} {Phys. Rev. A}\ }\textbf {\bibinfo {volume}
  {97}},\ \bibinfo {pages} {062124} (\bibinfo {year} {2018})}\BibitemShut
  {NoStop}%
\bibitem [{\citenamefont {Mitchison}\ and\ \citenamefont
  {Plenio}(2018)}]{Mitchison2018}%
  \BibitemOpen
  \bibfield  {author} {\bibinfo {author} {\bibfnamefont {M.~T.}\ \bibnamefont
  {Mitchison}}\ and\ \bibinfo {author} {\bibfnamefont {M.~B.}\ \bibnamefont
  {Plenio}},\ }\href {http://stacks.iop.org/1367-2630/20/i=3/a=033005}
  {\bibfield  {journal} {\bibinfo  {journal} {New J. Phys.}\ }\textbf {\bibinfo
  {volume} {20}},\ \bibinfo {pages} {033005} (\bibinfo {year}
  {2018})}\BibitemShut {NoStop}%
\bibitem [{\citenamefont {Hofer}\ \emph {et~al.}(2017)\citenamefont {Hofer},
  \citenamefont {Perarnau-Llobet}, \citenamefont {Miranda}, \citenamefont
  {Haack}, \citenamefont {Silva}, \citenamefont {Brask},\ and\ \citenamefont
  {Brunner}}]{1367-2630-19-12-123037}%
  \BibitemOpen
  \bibfield  {author} {\bibinfo {author} {\bibfnamefont {P.~P.}\ \bibnamefont
  {Hofer}}, \bibinfo {author} {\bibfnamefont {M.}~\bibnamefont
  {Perarnau-Llobet}}, \bibinfo {author} {\bibfnamefont {L.~D.~M.}\ \bibnamefont
  {Miranda}}, \bibinfo {author} {\bibfnamefont {G.}~\bibnamefont {Haack}},
  \bibinfo {author} {\bibfnamefont {R.}~\bibnamefont {Silva}}, \bibinfo
  {author} {\bibfnamefont {J.~B.}\ \bibnamefont {Brask}}, \ and\ \bibinfo
  {author} {\bibfnamefont {N.}~\bibnamefont {Brunner}},\ }\href@noop {}
  {\bibfield  {journal} {\bibinfo  {journal} {New J. Phys.}\ }\textbf {\bibinfo
  {volume} {19}},\ \bibinfo {pages} {123037} (\bibinfo {year}
  {2017})}\BibitemShut {NoStop}%
\bibitem [{\citenamefont {González}\ \emph {et~al.}(2017)\citenamefont
  {González}, \citenamefont {Correa}, \citenamefont {Nocerino}, \citenamefont
  {Palao}, \citenamefont {Alonso},\ and\ \citenamefont
  {Adesso}}]{doi:10.1142/S1230161217400108}%
  \BibitemOpen
  \bibfield  {author} {\bibinfo {author} {\bibfnamefont {J.~O.}\ \bibnamefont
  {González}}, \bibinfo {author} {\bibfnamefont {L.~A.}\ \bibnamefont
  {Correa}}, \bibinfo {author} {\bibfnamefont {G.}~\bibnamefont {Nocerino}},
  \bibinfo {author} {\bibfnamefont {J.~P.}\ \bibnamefont {Palao}}, \bibinfo
  {author} {\bibfnamefont {D.}~\bibnamefont {Alonso}}, \ and\ \bibinfo {author}
  {\bibfnamefont {G.}~\bibnamefont {Adesso}},\ }\href {\doibase
  10.1142/S1230161217400108} {\bibfield  {journal} {\bibinfo  {journal} {Open
  Syst. Inf. Dyn.}\ }\textbf {\bibinfo {volume} {24}},\ \bibinfo {pages}
  {1740010} (\bibinfo {year} {2017})}\BibitemShut {NoStop}%
\bibitem [{\citenamefont {Ol\v{s}ina}\ \emph {et~al.}()\citenamefont
  {Ol\v{s}ina}, \citenamefont {Dijkstra}, \citenamefont {Wang},\ and\
  \citenamefont {Cao}}]{Dijkstra14arxiv}%
  \BibitemOpen
  \bibfield  {author} {\bibinfo {author} {\bibfnamefont {J.}~\bibnamefont
  {Ol\v{s}ina}}, \bibinfo {author} {\bibfnamefont {A.~G.}\ \bibnamefont
  {Dijkstra}}, \bibinfo {author} {\bibfnamefont {C.}~\bibnamefont {Wang}}, \
  and\ \bibinfo {author} {\bibfnamefont {J.}~\bibnamefont {Cao}},\ }\href@noop
  {} {\bibinfo  {journal} {arXiv:1408.5385}\ }\BibitemShut {NoStop}%
\bibitem [{\citenamefont {del Pino}\ \emph {et~al.}(2018)\citenamefont {del
  Pino}, \citenamefont {Schr\"oder}, \citenamefont {Chin}, \citenamefont
  {Feist},\ and\ \citenamefont {Garcia-Vidal}}]{delpino18}%
  \BibitemOpen
\bibfield  {journal} {  }\bibfield  {author} {\bibinfo {author} {\bibfnamefont
  {J.}~\bibnamefont {del Pino}}, \bibinfo {author} {\bibfnamefont {F.~A.
  Y.~N.}\ \bibnamefont {Schr\"oder}}, \bibinfo {author} {\bibfnamefont {A.~W.}\
  \bibnamefont {Chin}}, \bibinfo {author} {\bibfnamefont {J.}~\bibnamefont
  {Feist}}, \ and\ \bibinfo {author} {\bibfnamefont {F.~J.}\ \bibnamefont
  {Garcia-Vidal}},\ }\href@noop {} {\bibfield  {journal} {\bibinfo  {journal}
  {Phys. Rev. Lett.}\ }\textbf {\bibinfo {volume} {121}},\ \bibinfo {pages}
  {227401} (\bibinfo {year} {2018})}\BibitemShut {NoStop}%
\bibitem [{\citenamefont {Garg}\ \emph {et~al.}(1985)\citenamefont {Garg},
  \citenamefont {Onuchic},\ and\ \citenamefont {Ambegaokar}}]{Garg1985}%
  \BibitemOpen
  \bibfield  {author} {\bibinfo {author} {\bibfnamefont {A.}~\bibnamefont
  {Garg}}, \bibinfo {author} {\bibfnamefont {J.~N.}\ \bibnamefont {Onuchic}}, \
  and\ \bibinfo {author} {\bibfnamefont {V.}~\bibnamefont {Ambegaokar}},\
  }\href {\doibase 10.1063/1.449017} {\bibfield  {journal} {\bibinfo  {journal}
  {J. Chem. Phys.}\ }\textbf {\bibinfo {volume} {83}},\ \bibinfo {pages} {4491}
  (\bibinfo {year} {1985})}\BibitemShut {NoStop}%
\bibitem [{\citenamefont {Thoss}\ \emph {et~al.}(2001)\citenamefont {Thoss},
  \citenamefont {Wang},\ and\ \citenamefont {Miller}}]{Thoss2001}%
  \BibitemOpen
  \bibfield  {author} {\bibinfo {author} {\bibfnamefont {M.}~\bibnamefont
  {Thoss}}, \bibinfo {author} {\bibfnamefont {H.}~\bibnamefont {Wang}}, \ and\
  \bibinfo {author} {\bibfnamefont {W.~H.}\ \bibnamefont {Miller}},\ }\href
  {\doibase 10.1063/1.1385562} {\bibfield  {journal} {\bibinfo  {journal} {J.
  Chem. Phys.}\ }\textbf {\bibinfo {volume} {115}},\ \bibinfo {pages} {2991}
  (\bibinfo {year} {2001})}\BibitemShut {NoStop}%
\bibitem [{\citenamefont {Carmichael}(2009)}]{carmichael2009statistical}%
  \BibitemOpen
  \bibfield  {author} {\bibinfo {author} {\bibfnamefont {H.~J.}\ \bibnamefont
  {Carmichael}},\ }\href@noop {} {\emph {\bibinfo {title} {Statistical Methods
  in Quantum Optics 2:~Non-Classical Fields}}}\ (\bibinfo  {publisher}
  {Springer},\ \bibinfo {year} {2009})\BibitemShut {NoStop}%
\bibitem [{\citenamefont {Dijkstra}\ and\ \citenamefont
  {Tanimura}(2015)}]{Dijkstra15}%
  \BibitemOpen
  \bibfield  {author} {\bibinfo {author} {\bibfnamefont {A.~G.}\ \bibnamefont
  {Dijkstra}}\ and\ \bibinfo {author} {\bibfnamefont {Y.}~\bibnamefont
  {Tanimura}},\ }\href {\doibase 10.1063/1.4917025} {\bibfield  {journal}
  {\bibinfo  {journal} {J. Chem. Phys.}\ }\textbf {\bibinfo {volume} {142}},\
  \bibinfo {pages} {212423} (\bibinfo {year} {2015})}\BibitemShut {NoStop}%
\bibitem [{\citenamefont {Loudon}(2000)}]{loudon2000quantum}%
  \BibitemOpen
  \bibfield  {author} {\bibinfo {author} {\bibfnamefont {R.}~\bibnamefont
  {Loudon}},\ }\href@noop {} {\emph {\bibinfo {title} {The Quantum Theory of
  Light}}}\ (\bibinfo  {publisher} {Oxford University Press},\ \bibinfo {year}
  {2000})\BibitemShut {NoStop}%
\bibitem [{\citenamefont {Breuer}\ and\ \citenamefont
  {Petruccione}(2002)}]{breuer2002theory}%
  \BibitemOpen
  \bibfield  {author} {\bibinfo {author} {\bibfnamefont {H.-P.}\ \bibnamefont
  {Breuer}}\ and\ \bibinfo {author} {\bibfnamefont {F.}~\bibnamefont
  {Petruccione}},\ }\href@noop {} {\emph {\bibinfo {title} {The Theory of Open
  Quantum Systems}}}\ (\bibinfo  {publisher} {Oxford University Press},\
  \bibinfo {year} {2002})\BibitemShut {NoStop}%
\bibitem [{\citenamefont {Agarwal}(2013)}]{agarwalQO}%
  \BibitemOpen
  \bibfield  {author} {\bibinfo {author} {\bibfnamefont {G.~S.}\ \bibnamefont
  {Agarwal}},\ }\href@noop {} {\emph {\bibinfo {title} {Quantum Optics}}}\
  (\bibinfo  {publisher} {Cambridge University Press},\ \bibinfo {year}
  {2013})\BibitemShut {NoStop}%
\bibitem [{Note1()}]{Note1}%
  \BibitemOpen
  \bibinfo {note} {For further details see
  the Supplemental Material}\BibitemShut {NoStop}%
\end{thebibliography}

\end{appendix}

\end{document}